\documentclass[useAMS,usenatbib,fleqn]{mnras}
\usepackage{amsmath}
\usepackage{multirow}
\usepackage{graphicx}
\usepackage{epsfig}
\usepackage{amssymb,amsmath}
\usepackage{aas_macros}
\usepackage{fixltx2e}
\title[MBHB TDEs]{
Tidal Disruption Events by a Massive Black Hole Binary
}

\author[Ricarte et al.]{Angelo Ricarte$^1$,
Priyamvada Natarajan$^1$,
Lixin Dai$^2$,
Paolo Coppi$^1$ \\
$^1$ Department of Astronomy, Yale University, 52 Hillhouse Avenue, New Haven, CT 06511 \\
$^2$ Department of Physics and Joint Space-Science Institute, University of Maryland, College Park, MD, 20742
}

\date{\today}

\begin{document}
\pagerange{\pageref{firstpage}--\pageref{lastpage}} \pubyear{2015}
\maketitle

\begin{abstract}

Massive black hole binaries (MBHBs) are a natural byproduct of galaxy mergers.  Previous studies have shown that flares from stellar tidal disruption events (TDEs) are modified by the presence of a secondary perturber, causing interruptions in the light curve.  We study the dynamics of TDE debris in the presence of a milliparsec-separated MBHB by integrating ballistic particle orbits in the time-varying potential of the binary.  We find that gaps in the light curve appear when material misses the accretion radius on its first return to pericentre.  Subsequent recurrences can be decomposed into ``continuous'' and ``delayed'' components, which exhibit different behaviour.  We find that this potential can substantially alter the locations of stream self-intersections.  When debris is confined to the plane, we find that close encounters with the secondary BH leave noticeable signatures on the fallback rate and can result in significant accretion onto the secondary BH.  Tight, equal-mass MBHBs accrete equally, periodically trading the infalling stream.  

\end{abstract}

\section{Introduction}

A tidal disruption event (TDE) occurs when a star ventures close enough to a black hole (BH) for its tidal forces to overcome the star's self-gravity \citep{Hills1975,Young+1977,Frank1978,Lacy+1982}.  During a TDE, stellar debris is elongated into a thin stream \citep{Kochanek1994,Guillochon+2014}, and its subsequent accretion manifests observationally as a flare in the UV/soft X-rays \citep{Rees1988} or optical wavebands \citep{Loeb&Ulmer1997,Ulmer+1998,Ulmer1999,Strubbe&Quataert2009}.  Its decay occurs on timescales of months to years, and tends to follow a characteristic $t^{-5/3}$ power law \citep{Rees1988,Phinney1989,Evans&Kochanek1989}.  Only BHs with masses $\lesssim 10^8 \, M_\odot$ can disrupt a solar-type star.  Above this cutoff, $r_t$ recedes beneath the Schwarzschild radius and stars are swallowed whole, unless the BH is rotating very rapidly \citep{Kesden2012}.  

An individual galaxy's central SMBH causes a TDE only once every $\sim10^{-5}$ years, extrapolating from X-ray, UV, and optical surveys \citep{Donley+2002,Gezari+2008,vanVelzen&Farrar2014}, although theoretically estimated rates tend to exceed observational ones by an order of magnitude \citep{Magorrian&Tremaine1999,Wang&Merritt2004,Alexander2012,Stone&Metzger2014}.  In any case, such rates are too infrequent for TDEs to contribute significantly to the growth of a SMBH.  However, TDEs are useful as observational probes due to their ability to temporarily illuminate the inactive population of MBHs.  They are especially critical at redshifts $z>0.1$ or BH masses $M_\bullet < 10^6 \, \mathrm{M}_\odot$, where dynamical mass estimates become challenging \citep{McConnell&Ma2013,Kormendy&Ho2013}.  Hence, TDEs will provide the strongest constraints on the low-mass end of the black hole mass function \citep{Stone&Metzger2014,MacLeod+2015}, which will decisively discriminate between different models of MBH seeding \citep[e.g., review by][]{Natarajan2011}.  So far, time domain surveys have detected $\sim 30$ TDEs \citep[review by][]{Komossa2015}, and future all-sky surveys such as the Large Synoptic Survey Telescope (LSST) are expected to observe thousands more \citep{Strubbe&Quataert2009}.  

TDEs can also provide a unique probe into the population of massive black hole binaries (MBHBs), which are created as a consequence of galaxy mergers \citep{Begelman+1980,Volonteri+2003}.  These binaries are thought to form as a result of a concert of dynamical processes on different scales.  In the first phase, dynamical friction degrades the orbit of an infalling BH and drives it to the centre of the galaxy potential \citep{Chandrasekhar1943}.  Below approximately 1 parsec, dynamical friction loses efficiency (the ``last parsec problem''), and other processes such as gas dynamics and stellar scattering are required to bring the BHs close enough to eventually merge via gravitational radiation \citep[see reviews by][]{Colpi2014,Volonteri+2015}.  Identifying and characterising the population of MBHBs would be paramount to testing our theories of dynamics and MBH hole formation.  

Contemporary techniques to identify candidates, such as searching for spatially-resolved  ``dual'' or ``offset'' active galactic nuclei (AGN), double-peaked emission lines, or periodic variability \citep{Comerford+2009,Komossa&Zensus2015}, are usually limited to systems where both BHs are accreting.  As we shall show, TDEs allow us to probe MBHBs which are non-accreting, rendering them undetectable by these aforementioned methods.  Fortuitously, recent studies have demonstrated that tidal disruption rates by a MBHB can be several orders of magnitude higher than those of a single MBH \citep{Chen+2009,Liu&Chen2013,Li+2015,LiG+2015}.  This does depend on the separation, however;  as a binary hardens, its TDE rate may dip below the single MBH rate \citep{Chen+2008} or level off to normal levels \citep{Li+2015}.  It is possible that this enhancement may help explain the recent finding that TDEs are preferentially hosted by post-merger galaxies \citep{Arcavi+2014}.  

It is when MBHBs shrink to milliparsec separations that the dynamics of a TDE are modified substantially by the presence of the binary companion.  These tight systems have periods on the order of years, coinciding with the decay timescale of a typical TDE flare.  \citet{Liu+2009} (henceforth L09) performed the first study of TDE light curves in the presence of a MBHB.  They found that some of them material does not always reach the central BH as quickly as it would around a single MBH, causing the canonical $t^{-5/3}$ flare to be truncated.  Following this truncation, discrete ``accretion islands'' of episodic accretion occur, although the physical origin of these islands has been left unexplored.  These predictions were fulfilled five years later, leading to the identification of the first milliparsec-separated MBHB candidate \citep{Liu+2014} (henceforth L14).  If more MBHBs are identified in this way, they will be pivotal for constraining models of BH growth and dynamics.

In this study, we take a closer look at the fallback of the stellar debris in the presence of a MBHB.  We extend the work of L09 and L14, seeking the physical understanding of their reported interruptions and ``accretion islands."  To do so, we simulate the fallback of tidal debris in the presence of a MBHB using a Runge-Kutta-Fehlberg orbital integrator.  We explore a range of free parameters: the binary mass ratio $q \equiv M_2/M_1$ where $M_1 \geq M_2$, the semimajor axis of the binary, $a$, and the position of the star around the MBH, given by $\phi$, the longitude of periapsis, and $\theta$, where $90^\circ-\theta$ is the inclination of the stellar orbit with respect to the plane of the MBHB.  For each run, we record the fallback rate of particles onto each black hole, and create visualisations of the system at 100 times throughout the simulation.  This allows us to decompose fallback into two distinct components, unpack the gravitational effects of the secondary BH, and locate stream intersections, all of which informs future hydrodynamical work.

As in L09 and L14, we elect not to include hydrodynamics and instead follow non-interacting particles on ballistic trajectories.  In general, hydrodynamical simulations of TDEs are challenging due to the wide dynamic range in both length and time scales that are required.  These studies have focused on single BHs rather than binaries.  At present, most state-of-the-art studies restrict themselves to special, rare cases to improve computational efficiency.  \citet{Guillochon+2014} and \citet{Shiokawa+2015} simulate the disruption of a white dwarf around an intermediate mass black hole to reduce the dynamic range.  Alternatively, \citet{Hayasaki+2015} and \citet{Bonnerot+2015} simulate the disruption of stars with relatively low eccentricity, where the maximum energies are lower.  \citet{Coughlin&Nixon2015} simulate the disruption of a solar mass star around a $10^6 \, \mathrm{M}_\odot$ BH with hydrodynamics and self-gravity, but specify an accretion radius instead of capturing circularization processes.  Extending the dynamic range to include the separation and period of a milliparsec MBHB would further increase computational cost.  Without hydrodynamics, we cannot capture the physics of shocks, which are critical for understanding circularisation processes.  Here, we rely on the guidance of hydrodynamical simulations of single BH TDEs, which capture stream's nozzle shock at pericentre passage \citep{Guillochon+2014} and stream self-intersections \citep{Shiokawa+2015,Hayasaki+2015,Bonnerot+2015}.  On the other hand, we can rapidly traverse a wide variety of parameters to guide the future hydrodynamical studies that will.  

This paper is organised as follows.  In \S\ref{sec:methods}, we explain our model for the stellar debris as well as numerical techniques.  In \S\ref{sec:results}, we first describe the main results of this study.  We first introduce the typical TDE with a MBHB, revealing that the system quickly grows cluttered with unaccreted material.  We then distinguish between continuous and delayed accretion, which leave different signatures in the light curves of these TDEs.  We finish this section with a study of TDEs in the plane, where the influence of the secondary BH is maximised.  In \S\ref{sec:special}, we explore a few special investigations: the timing of first truncation, sensitivity to the accretion radius we specify, and the locations of stream crossings.  In \S\ref{sec:discussion} we consolidate our parameter studies and discuss final caveats of the fallback rate curves we generate.  In \S\ref{sec:conclusions}, we summarise and highlight the main conclusions of this study.

\section{Methods}
\label{sec:methods}

\subsection{Physical Model}
\label{sec:tde_physics}

In this work, we consider the complete disruption of a solar-type star on a parabolic orbit, since the two-body scattering mechanism that generates disrupted stars tends to produce stars from large distances \citep{Wang&Merritt2004}.  We disrupt stars only around the primary BH, as the probability of stellar disruption by the secondary BH is low if the BHs have very unequal masses \citep{Chen+2008,Chen+2009}.  For simplicity, the MBHBs we consider are on circular orbits.  Our simulations begin with a star at pericentre at the tidal disruption radius, given by

\begin{equation}
r_t \equiv R_* \left( \frac{M_\bullet}{M_*} \right)^{1/3} \label{eqn:r_t}
\end{equation}

\noindent where $R_*$ is the radius of the star, $M_*$ is the mass of the star, and $M_\bullet$ is the mass of the black hole.  In TDE parlance, this orbit corresponds to a penetration parameter $\beta \equiv r_t/r_p = 1$.   Hydrodynamical simulations show that varying $\beta$ for a single BH TDE can change the internal structure of a stream \citep{Guillochon&Ramirez-Ruiz2013}, but this has no impact on our results other than to change the amplitude and slope of underlying fallback rate curves.

When a star is disrupted, particles are distributed in specific energy between the least bound orbits, $E_\mathrm{lb} = E_t + \Delta E$, and the most bound orbits, $E_\mathrm{mb} = E_t - \Delta E$ \citep{Rees1988}.  Here, $\Delta E$ is the spread in specific energies, and $E_t$ is the specific binding energy of the star to the black hole, which is zero for a parabolic orbit.  In the ``freezing model," $\Delta E$ is given exactly by the difference in energy due to the potential across the stellar radius, $R_*$.  Usually, by Taylor expanding a {\it Newtonian} potential at $r_t$, $\Delta E$ is given by

\begin{equation}
\Delta E = \frac{k G M_\bullet R_*}{r_t^2} \label{eqn:DeltaE}
\end{equation}

\noindent where the extra factor of $k \in [1,3]$ has been inserted to account for the possible spin-up of a star as it is disrupted.  For consistency with L09, we set $k=2.5$ as a compromise between the two cases.  This choice has little impact on our results: increasing (decreasing) $k$ only increases (decreases) (i) the amount of bound material that is immediately accreted most rapidly, with minimal interaction with the MBHB, and (ii) the amount of unbound material that is ejected at the greatest speeds, which is likely to escape the system without interacting with either black hole.  Under the assumption that mass is distributed equally across energy bins \citep{Rees1988,Phinney1989}, the fallback rate is given by

\begin{align}
\frac{dM}{dt} &= \frac{M_*}{2 \Delta E} \cdot \frac{(2 \pi G M_\bullet)^{2/3}}{3} t^{-5/3} \\
&= \frac{M_*}{3 t_\mathrm{min}}\left( \frac{t}{t_\mathrm{min}} \right)^{-5/3} \label{eqn:dMdt}
\end{align}

\noindent for $t > t_\mathrm{min}$, where $M_*$ is the mass of the star and $t_\mathrm{min}$ is the time it takes for the most bound material to fall back onto the black hole, given via Kepler's 3rd law by

\begin{equation}
t_\mathrm{min} = 2 \pi G M_\bullet (2 \Delta E)^{-3/2} \label{eqn:t_min}
\end{equation}

\noindent Constant $dM/dE$ finds support in hydrodynamical simulations \citep[e.g.,][]{Evans&Kochanek1989}. It may be modified, however, due to stellar structure \citep{Lodato+2009,Guillochon&Ramirez-Ruiz2013}.  We neglect these effects, which would alter the shape of the underlying fallback rate curves, but not the effects of the binary that we explore.

The addition of a secondary BH introduces another characteristic timescale, the period of the binary, $T_\bullet$.  L09 use the dynamical stability analysis of \citet{Mardling&Aarseth2001} to predict that the fallback of equation \ref{eqn:dMdt} should be ``truncated'' at time

\begin{equation}
T_\mathrm{tr} \approx \frac{T_\bullet}{4.7}(1+q)^{-1/10}\frac{(1-e)^{9/5}}{(1+e)^{3/5}}(1-0.3i/180^\circ)^{-3/2} \label{eqn:T_tr}
\end{equation}

\noindent where $q \equiv M_2/M_1$ is the mass ratio of the MBHB, $e$ is the eccentricity of the MBHB, and $i$ is the inclination of the plane of the stellar orbit with respect to that of the MBHB.  We specify the initial velocity of the stellar material such that $i = 90^\circ - \theta$.  Truncation results in subsequent ``accretion islands,'' periods of accretion that appear on binary timescales.  Since all of our binaries have $e=0$, this reduces to $T_\mathrm{tr} \approx T_\bullet/4.7\cdot(1+q)^{-1/10}(1-0.3i/180^\circ)^{-3/2}$.  

\subsection{Numerical Techniques}

Individual particle orbits are integrated using an adaptive order 7(8) Runge-Kutta-Fehlberg algorithm \citep[see table 5.3]{Hairer+1993}, where the eighth-order estimate is kept (``local extrapolation'').  Rather than being treated as particles, the two orbiting black holes are used to define a time-varying potential.  We test for convergence by extracting fallback rate curves from a simulation with significant chaos, with parameters (as defined below) given by $M_\bullet = 10^7 \, M_\odot$, $q=1$, $a = 1$ milliparsec, $\theta = \pi/4$, and $\phi=-5 \pi/6$.  The tolerated fractional error per time step is varied between $10^{-12}$, $10^{-13}$, and $10^{-14}$.  We notice no differences between the fallback rates as the tolerance is modified, and conservatively set it to $10^{-13}$ in our simulations.

In our simulations, the tidal radius is only $\sim 10$ times the gravitational radius for $10^7 \, \mathrm{M_\odot}$ MBHs, and $\sim 50$ times the gravitational radius for $10^6 \, \mathrm{M_\odot}$ MBHs, where proximity to the gravitational radius, $r_g =  GM_\bullet/c^2$, implies significant effects due to general relativity (GR).  It is therefore important to use a potential that captures at least some relativistic effects.  While the previous work of L09 and L14 employs the potential of \citet{Paczynsky&Wiita1980}, we elect to use the potential of \citet{Wegg2012}.  An improvement over the more familiar potential of \citet{Paczynsky&Wiita1980}, this potential is designed for nearly parabolic orbits to reproduce the apsidal precession with greater accuracy.  It is given by

\begin{equation}
U(r) = -\frac{GM_\bullet}{r}\left[ \alpha + \frac{1-\alpha}{1 - R_x/r} + \frac{R_y}{r}\right] \label{eqn:WeggC}
\end{equation}

\noindent where the \citet{Paczynsky&Wiita1980} potential is recovered by setting the three free parameters to $\alpha=0$, $R_x = 2GM_\bullet/c^2$, and $R_y=0$.  The potential we use (potential C) reproduces the apsidal precession angle with $<1$\% error.  It sets

\begin{align}
\alpha &= -\frac{4}{3} (2+\sqrt{6}) \\
R_x &= (4\sqrt{6} - 9) \frac{GM_\bullet}{c^2} \\
R_y &= -\frac{4}{3}(2\sqrt{6}-3) \frac{GM_\bullet}{c^2}
\end{align}

\noindent For self-consistency, we Taylor expand equation \ref{eqn:WeggC} at $r=r_t$, and replace the Newtonian estimate of $\Delta E$ in equation \ref{eqn:DeltaE} with 

\begin{equation}
\Delta E = \frac{k G M_\bullet R_*}{r_t^2} \left[ \alpha - \frac{1-\alpha}{(1 - R_x/r_t)^2} + \frac{2R_y}{r_t}\right] \label{eqn:DeltaE_Wegg}
\end{equation}

\noindent in our simulations.  We sample the specific binding energy range $E \in [-\Delta E,\Delta E]$ with $10^5$ particles using a hybrid logarithmic-linear scheme.  85\% of the particles are allocated to a logarithmic sampling between $[-\Delta E, E_\mathrm{max}]$, where the cutoff $E_\mathrm{max}$ is the specific energy of a bound particle that would be at apocentre at the end of the simulation in the absence of the secondary black hole.  If $T_\mathrm{max}$ represents the run-time of the simulation, then $E_\mathrm{max} = 0.5(\pi G M_\bullet / T_\mathrm{max})^{2/3}$.  Using $m$ of the $n$ total particles to sample this range, the $i$th particle has total energy 

\begin{equation}
E_i = \Delta E \left( \frac{E_\mathrm{max}}{\Delta E} \right)^{(i+0.5)/m} \label{eqn:itoE_log}
\end{equation}

\noindent and is assigned a mass $M_i = |0.5M_*E_i \ln(E_\mathrm{max}/\Delta E)/m\Delta E|$ to fix $dM/dE = M_*/2 \Delta E$. This sampling ensures that the stream is well-sampled, and that the mass fallback rate is resolved even at late times when orders of magnitude less mass should be infalling.  

With the addition of the binary companion, we notice that there are setups where otherwise unbound material can be accreted.  These events are rare and dependent on the initial conditions, occurring most obviously if the binary companion is high-mass and reaches the stream before any of its constituent particles have reached apocentre.  In order to represent some of the otherwise unbound material, but not spend too much computational time resolving it, the remaining 15\% of the particles are allocated to the energy range $E \in (E_\mathrm{max}, \Delta E]$.  This range is sampled linearly, with

\begin{equation}
E_i = E_\mathrm{max} + \left( \frac{i+0.5-m}{n-m} \right) \cdot(-\Delta E - E_\mathrm{max}) \label{eqn:itoE_lin}
\end{equation}

\noindent and $m_i = M_*/2\Delta E \cdot (\Delta E + E_\mathrm{max})/(n-m)$.

\begin{figure*}
\begin{center}
\includegraphics[width=\textwidth]{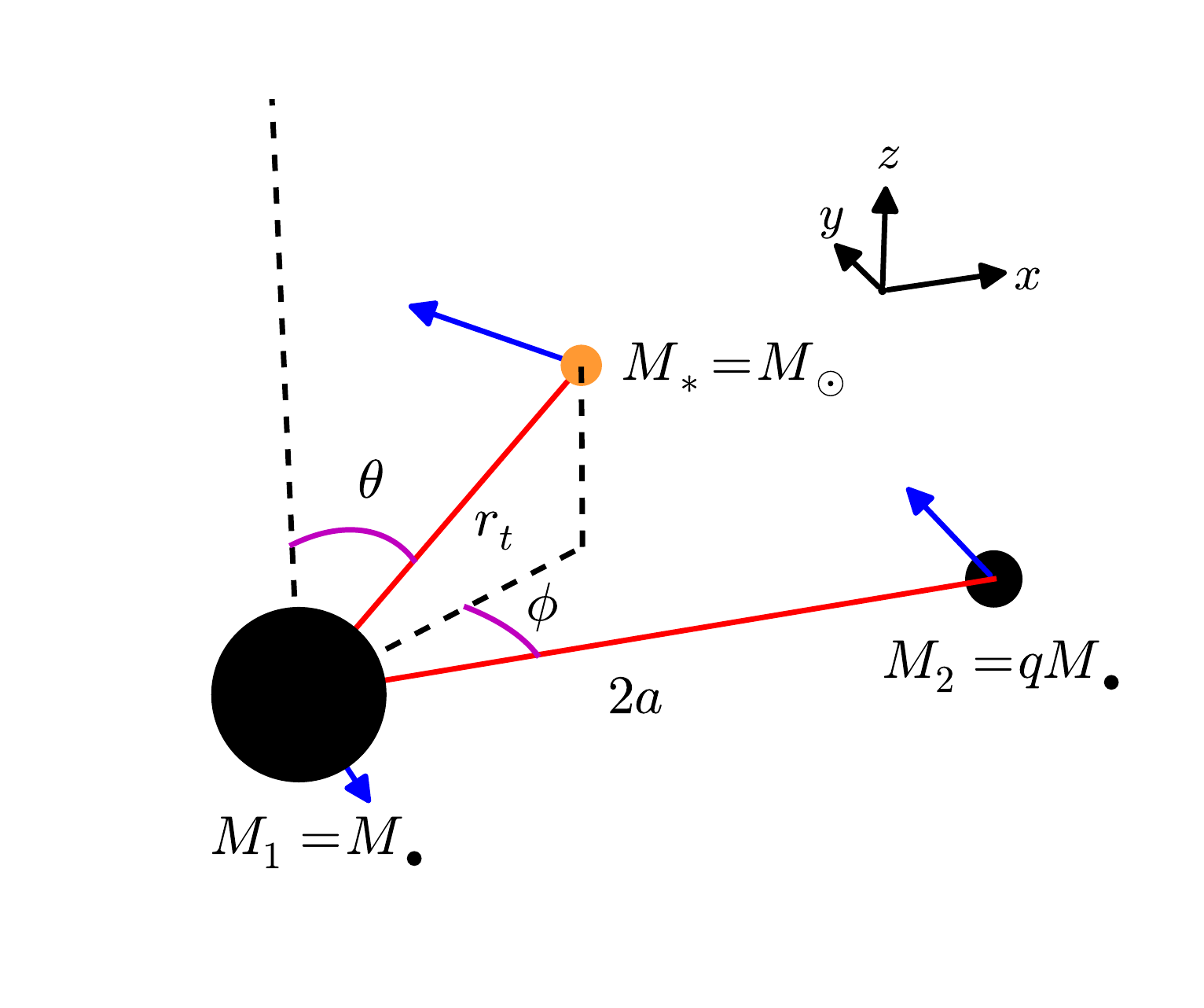}
\caption{An illustration of our setup.  The MBHB defines the $x-y$ plane.  The stellar debris is initialised at pericentre a distance $r_t$ from the star, with angular coordinates $\phi$ and $\theta$.  The two MBHs are in a circular orbit, separated by a distance $2a$.  The direction of each object's velocity is plotted with blue vectors.  This figure is schematic and not to scale.  \label{fig:visualizeSetup}}
\end{center}
\end{figure*}

In a single run, each particle is initialised with the same position in space, given by the radius $r_t$ from the primary black hole, and angular coordinates $\phi$ and $\theta$.  These are oriented such that the secondary BH at $t=0$ has coordinates $\phi=0^\circ$ and $\theta=90^\circ$.  Differences in energy are entirely given by differences in pericentric velocities.  For particle $i$, this is given by $v_i = \sqrt{2(U_t - E_i)}$, where $U_t$ is the specific potential energy at $r_t$ calculated using equation \ref{eqn:WeggC}.  The direction of the velocity vector is specified as a function of $\phi$, but independent of $\theta$, such that $i = 90^\circ - \theta$.  An illustration of our setup is provided in Figure \ref{fig:visualizeSetup} (not to scale).

Particles are removed if they (re-)enter the accretion radius of {\it either} BH, defined $r_\mathrm{acc} \equiv 2r_t$, where we assume debris material rapidly circularises and accretes in the same manner as in the presence of a single MBH, where orbital GR precession facilitates stream self-intersections \citep{Shiokawa+2015,Bonnerot+2015,Hayasaki+2015}.  Using the same radius for the secondary black hole (scaled to the smaller hole's mass) is arbitrary, but does not qualitatively alter our results.

The free parameters which we explore in this study are the binary mass ratio, $q$, the mass of the primary black hole, $M_\bullet$, the semimajor axis of the binary system, $a$, and the initial position of the star, given by $\phi$ and $\theta$.  Most of the results in this work follow from a grid exploring these parameters sampling every combination given by $q \in \{0.01,0.1,1\}$, $M_\bullet/\mathrm{M}_\odot \in \{10^6, 10^7\}$, $a/\mathrm{milliparsec} \in \{0.1, 1\}$, $\phi \in \{0,120^\circ,240^\circ\}$, and $\theta \in \{0^\circ,45^\circ,90^\circ\}$.  These are each integrated for 3 binary periods, to examine physically relevant timescales.  Positions and velocities are saved for 100 snapshots throughout each run, which are later used for visualisations, mass fallback rate curves, and other products.

\section{Results}
\label{sec:results}

\subsection{The Typical, Messy TDE with a MBHB}

During a TDE around a single MBH, the stellar debris falls back to pericentre according to equation \ref{eqn:dMdt}.  Assuming an efficient circularisation mechanism, such as shocks induced by apsidal precession, this material is promptly accreted and the MBH radiates with luminosity $L \propto \dot{M} \propto t^{-5/3}$.  In the case of a binary, L09 first discovered that material does not always reach the accretion radius as promptly as it would in single BH TDEs, causing gaps in the light curve.  Accretion may recur after the first interruption, but not necessarily in a manner that follows the canonical $t^{-5/3}$ fallback rate.  We seek an explanation of why some material avoids accretion, and what the ultimate fate of this material is.

\begin{table*}
   \centering
   \begin{tabular}{ ccccc }
   \hline
    $t_\mathrm{min}$ (yr) & $T_\bullet$ (yr) & $r_g$ (mpc) & $r_t$ (mpc) & $a_\mathrm{mb}$ (mpc)\\
    0.034 & 2.5 & $4.8 \times 10^{-4}$ & $4.9 \times 10^{-3}$ & 0.11 \\
    \hline
   \end{tabular}
   \caption{Physically relevant values for our typical MBHB TDE, where $a=1$ milliparsec and $M_\bullet=10^7 \, \mathrm{M}_\odot$.  $t_\mathrm{min}$ is the minimum return time of the stellar material (and also decay timescale of the flare), $T_\bullet$ is the period of the MBHB, $r_g$ is the gravitational radius of the primary BH, $r_t$ is the tidal radius of the primary BH, and $a_\mathrm{mb}$ is the semimajor axis of the most-bound stellar material.}
   \label{tab:numbers}
\end{table*}

In Figure \ref{fig:vis_typical_p1}, we visualise the results of a MBHB TDE that is given typical parameters:  $q=0.1$, $a=1$ milliparsec, $\phi=120^\circ$, $\theta = 45^\circ$, and $M_\bullet=10^7 \, \mathrm{M}_\odot$.  This visualisation is projected onto the plane of the binary, which defines the $xy$ plane.  Physically relevant values for this run are presented in table \ref{tab:numbers}.  In this visualisation, and those that follow, each particle is colour-coded according to its index, which maps one-to-one with initial energy by equations \ref{eqn:itoE_log} and \ref{eqn:itoE_lin}.  Red particles have the most positive energies, while purple particles have the most negative energies (and are thus usually accreted and removed from the simulation the fastest).  Note that particle and black hole markers are not to scale, although the relative sizes of the black hole markers accurately represent the mass ratio $q$.  In the top-left of each panel, $n$ denotes the number of the snapshot (out of 100), and $t$ represents the time since disruption.

\begin{figure*}
\centering
\includegraphics[width=\textwidth]{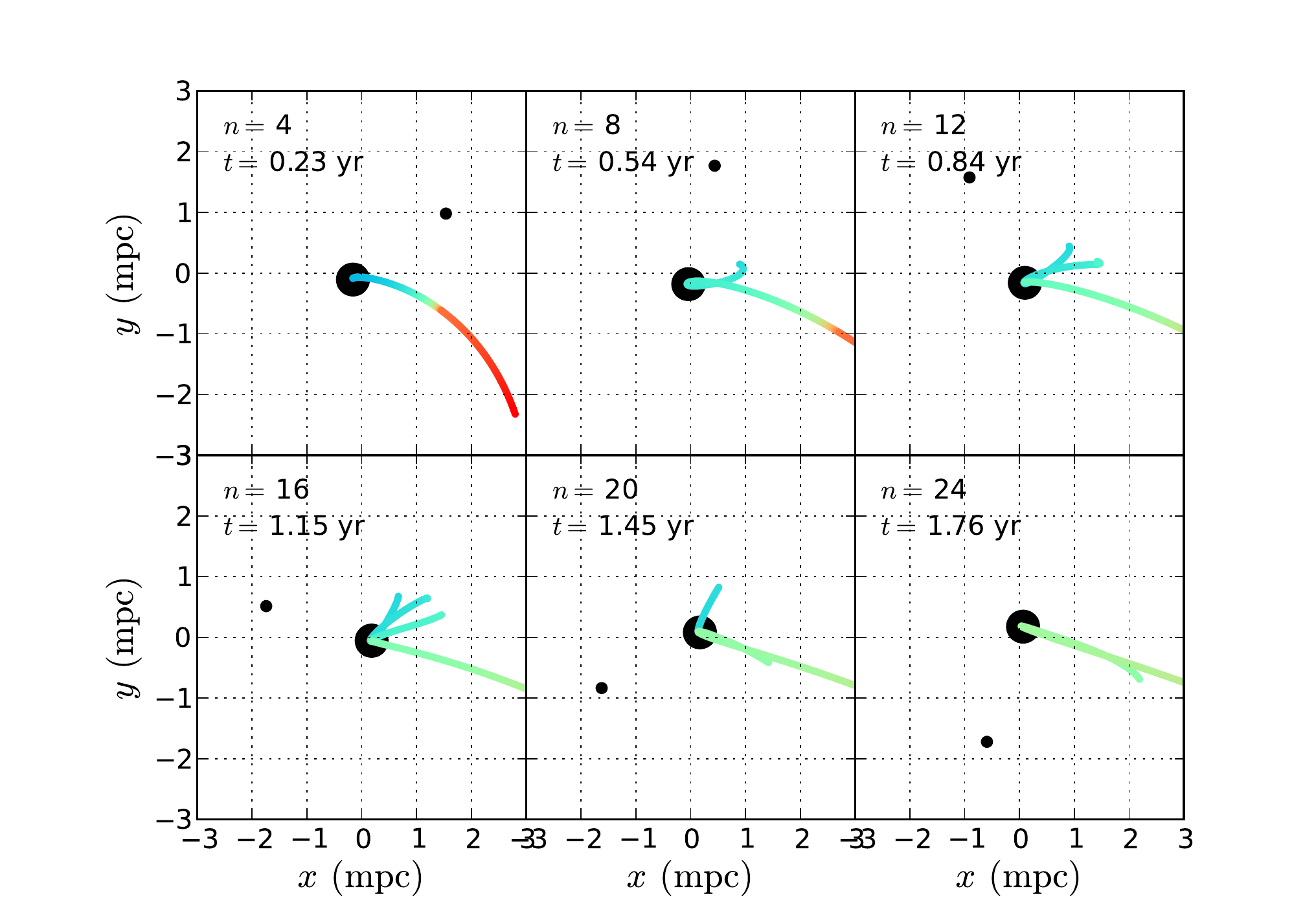}
\caption{Visualisation of a typical MBHB TDE.  Here, $q=0.1$, $a=1$ milliparsec, $\phi=120^\circ$, $\theta=45^\circ$, and $M_\bullet=10^7 \, \mathrm{M}_\odot$.  In this visualisation, and those to come, particles are colour-coded according to their initial energies, with purple being the most bound (already accreted by the first panel) and red being the least bound.  Particle and black hole markers are not to scale, although the relative sizes of the black hole markers accurately represent the mass ratio $q$.  $n$ denotes the number of each panel's snapshot, while $t$ represents the time since disruption.  In the second panel, the infalling stream misses the accretion radius and continues orbiting for a few periods until it is eventually accreted in the subsequent panels.  In the fifth and sixth panels, the infalling stream begins to miss again, indicating the beginning of another stage without accretion.  Note that this visualisation is a projection onto the MBHB plane, so overlaps do not necessarily imply intersection.  \label{fig:vis_typical_p1}}
\end{figure*}

For a typical MBHB TDE such as this one, the early stages of accretion proceed exactly as expected for a single MBH.  In the first panel of Figure \ref{fig:vis_typical_p1}, the most-bound particles have already been accreted, and the stream extends just as it would in the case of a single MBH.  The first interruption occurs in the second panel---particles have missed the accretion radius and have instead continued to orbit the central MBH.  Note that this occurs without any tidal stream-secondary BH close encounters, resulting purely from the gravitational potential of the MBHB.  In this particular simulation, the particles which initially miss accrete onto the MBH after a few extra orbits (by the 5th panel), but this is not always the case.  Then, in the 5th and 6th panels, new material begins to miss the primary BH, overlapping with the infalling stream.  Once the secondary BH reaches closest approach with the stream, the structure of the stream becomes much more complex.  We visualise the structure that evolves over the course of several additional periods in Figure \ref{fig:vis_typical_p2}.  More and more material misses the accretion radius and orbits within the system, and it is clear that shocks beyond the accretion radius that we do not capture may grow increasingly important.

\begin{figure*}
\centering
\includegraphics[width=\textwidth]{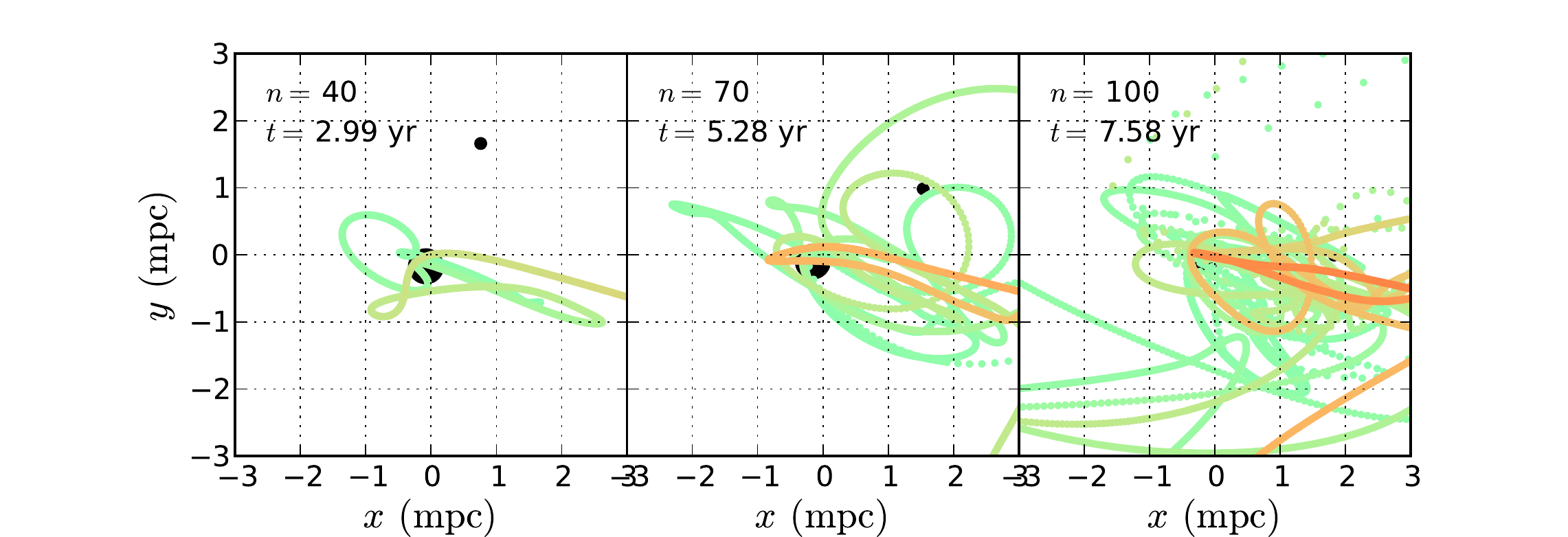}
\caption{Continuation of visualisation of a typical MBHB TDE, extended from Figure \ref{fig:vis_typical_p1}.  After closest approach between the secondary BH and the stream, the dynamics are significantly perturbed.  Shocks due to stream self-intersections that are not captured in our simulations should grow increasingly important as material accumulates that has not accreted onto either BH.  Note that this visualisation is a projection onto the MBHB plane, so overlaps do not necessarily imply intersection.  \label{fig:vis_typical_p2}}
\end{figure*}

\begin{figure*}
\centering
\includegraphics[width=0.5\textwidth]{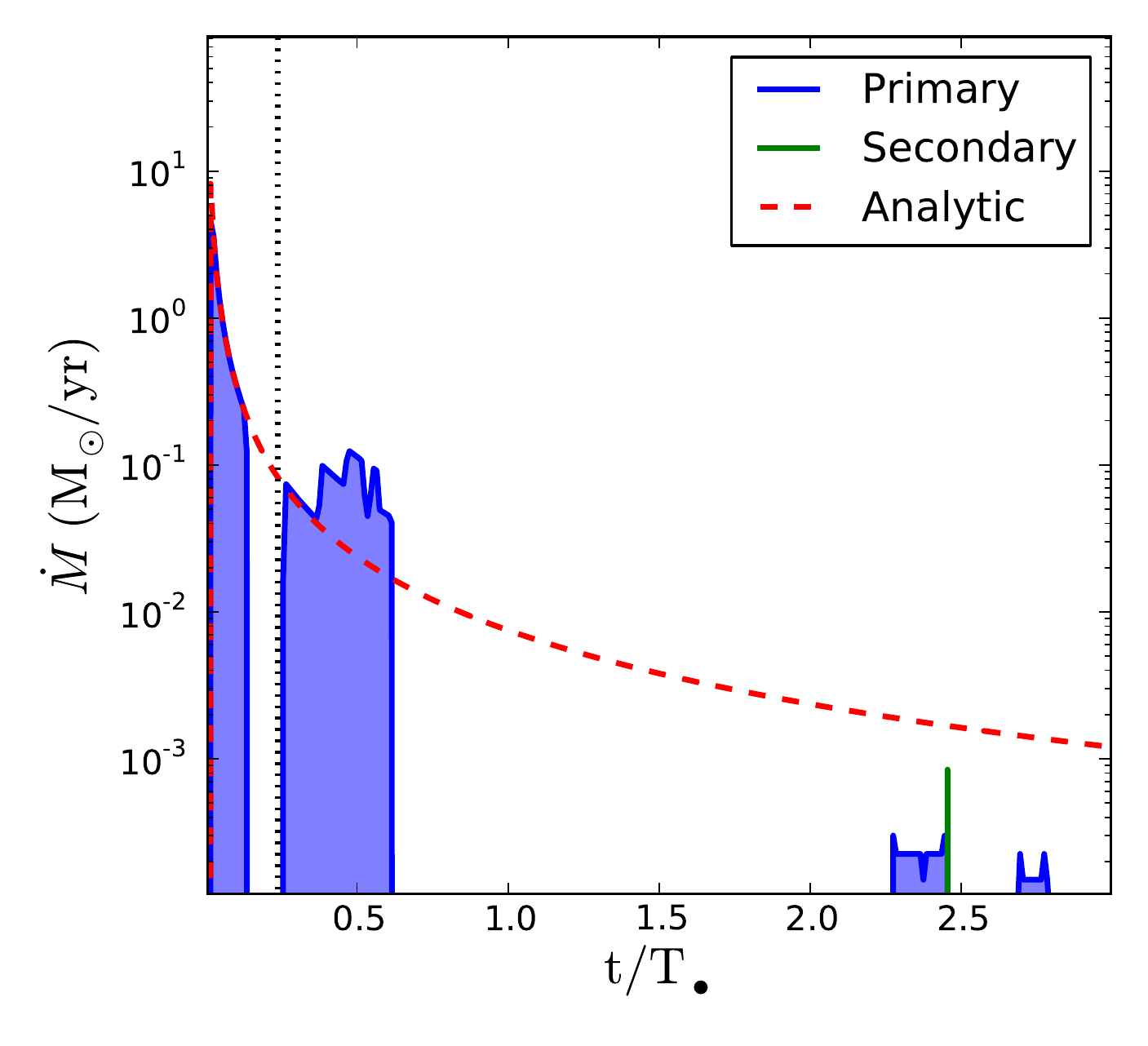}
\caption{Mass fallback rate for the simulation in which $q=0.1$, $a=1$ milliparsec, $\phi=120$, and $M_\bullet=10^7 \, \mathrm{M}_\odot$, which is shown in Figures \ref{fig:vis_typical_p1} and \ref{fig:vis_typical_p2}.  Blue and green curves correspond to accretion onto the primary and secondary black holes respectively.  For the case of a single black hole, the analytic solution of equation \ref{eqn:dMdt} is plotted as a red dashed line.  The dotted vertical black line marks the time at which truncation is predicted by equation \ref{eqn:T_tr}.  \label{fig:typicalFallback}}
\end{figure*}

This physical behaviour is imprinted into the mass fallback rate, shown in Figure \ref{fig:typicalFallback}.  In this figure, and those to come, fallback rate curves are generated from the particles that enter the accretion radius of either BH and are removed from the simulation.  The blue curves represent accretion onto the primary BH, while the green curves represent accretion onto the secondary BH.  In most cases, especially those out of the plane, the mass accreted by the secondary black hole is small.  The red dashed curve displays the analytic fallback rate given for a single MBH by equation \ref{eqn:dMdt}.  Finally, the dotted black vertical line represents the estimate for the truncation time given by equation \ref{eqn:T_tr}.

In this figure, the fallback rate onto the primary is matched perfectly until the first interruption.  As noted first by L09, the first interruption of accretion coincides roughly, but not perfectly, with the time predicted by equation \ref{eqn:T_tr}.  We have determined that gaps in the fallback rate result from material that misses the accretion radius, but is not usually involved in close encounters with the secondary MBH.  The first interruption in this particular run, occurring around $0.2 T_\bullet$, also leads to excess accretion around $0.5 T_\bullet$, when the missed material finally reaches the accretion radius.  After the second interruption, almost no accretion occurs in our simulations.  However, it is clear from Figure \ref{fig:vis_typical_p2} that shocks induced by stream self-intersections outside the accretion radius should provide an energy-loss mechanism that would increase the amount of accretion onto the primary BH.  It may therefore be too hasty to conclude that the dearth of accretion following the second interruption is real.  Shocks outside the accretion radius may increase the amount of material accreted, bringing the accretion radius prescription into question.  Hence, it is of interest for us to (i) identify accretion that is most likely to remain invariant under the addition of hydrodynamics and (ii) locate the shocks that should occur in a hydrodynamical simulation.

\subsection{Continuous vs. Delayed Accretion}
\label{ssec:continuousDelayed}

As introduced in the previous section, gaps in the mass fallback rate result in the accumulation of material orbiting the system that missed the primary BH.  Ultimately, this material might eventually leave the system or accrete onto either BH after a delay.  Meanwhile, even late in these simulations, there exists a long-lived coherent stream that continues to fall toward the primary BH.  In this section, we show that this behaviour allows us to decompose fallback rate curves into two components:  continuous accretion and delayed accretion.  Continuous accretion corresponds to material that accretes as expected by the canonical $t^{-5/3}$ power law, seemingly unperturbed by presence of the secondary BH.  Delayed accretion, on the other hand, originates from the interruptions of this power law accretion.  In the fallback rate, continuous accretion manifests as periodic blocks of accretion that follow the $t^{-5/3}$ power law, while delayed accretion contributes to additional spikes and peaks after interruptions that appear without a clear pattern.

We isolate these two modes in our simulations by making cuts according to the timing of each particle's accretion compared to the time it would have been accreted by a single BH.  We say that particle $i$ accretes {\it continuously} if the time of its accretion, $t_{\mathrm{acc},i}$, satisfies

\begin{equation}
t_{\mathrm{acc},i} \leq 1.5 \cdot T_i = 1.5 \cdot [2 \pi G M_\bullet (2 E_i)^{-3/2}] \label{eqn:continuous}
\end{equation}

\noindent That is, continuous accretion consists of those particles that accrete no more than 50\% later than the time, $T_i$, that it would take to accrete in the absence of the secondary BH.

\begin{figure*}
\begin{center}$
\begin{array}{ccc}
\includegraphics[width=0.33\textwidth]{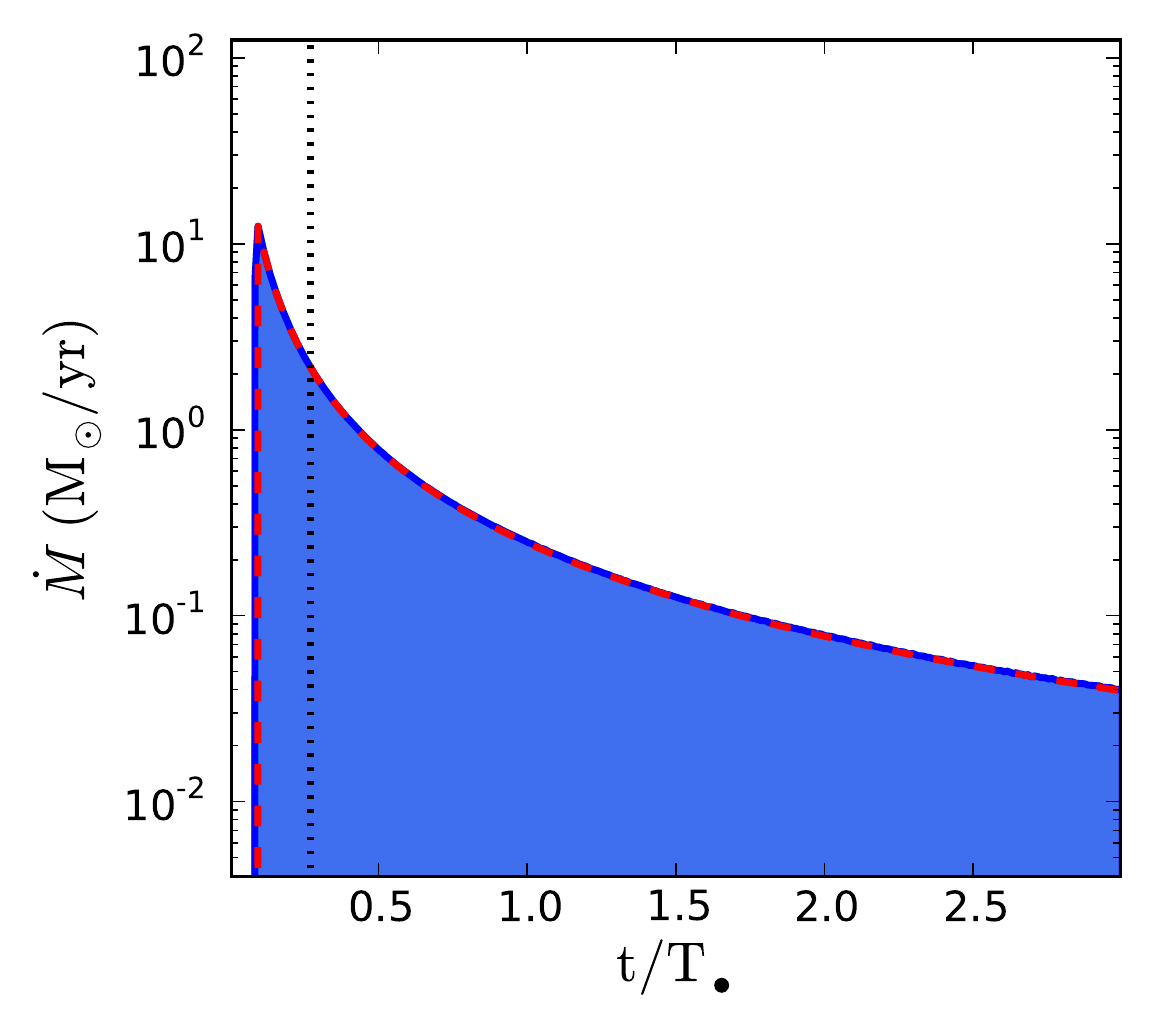} &
\hspace{-15pt} 
\includegraphics[width=0.33\textwidth]{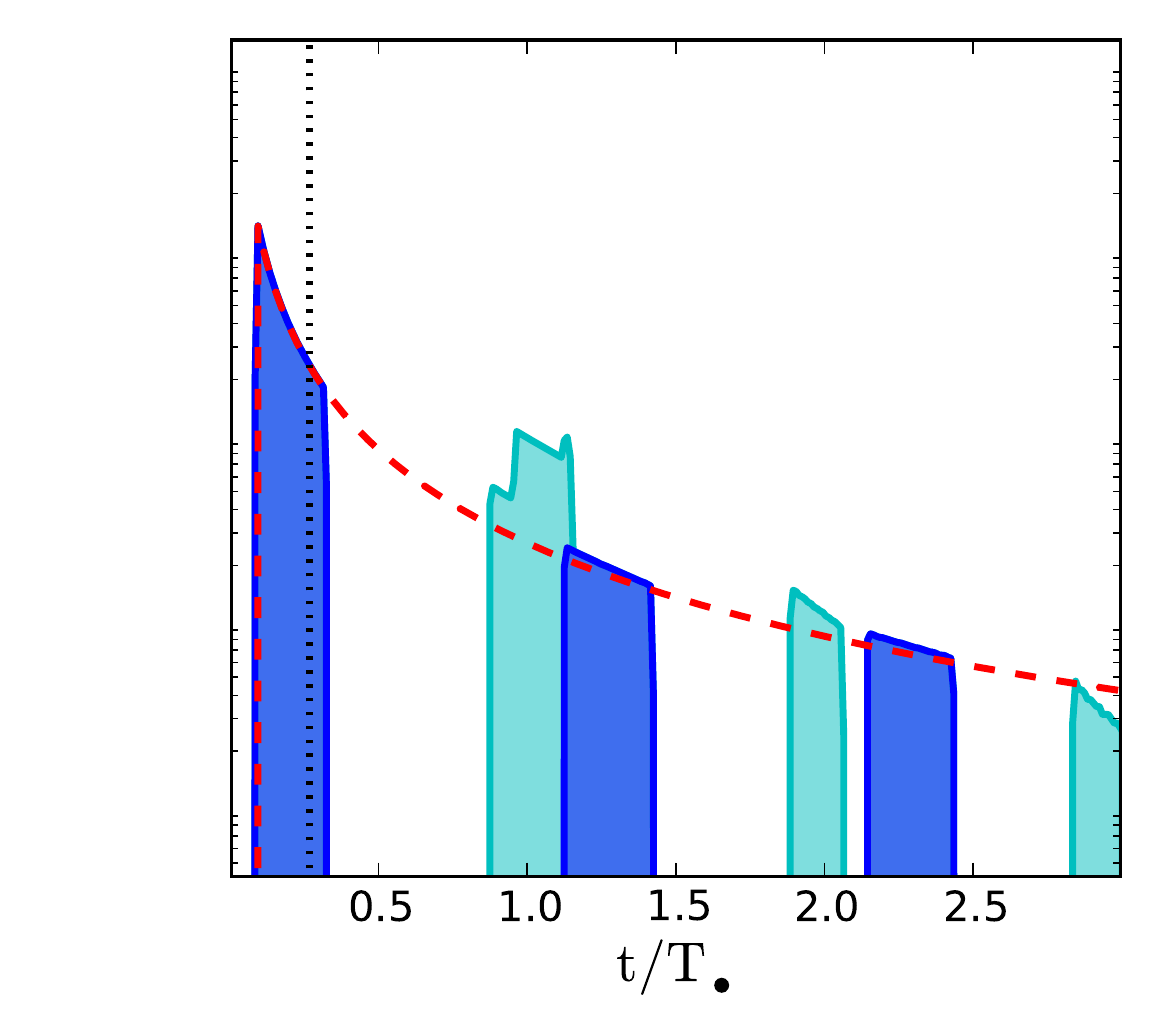} &
\hspace{-15pt}
\includegraphics[width=0.33\textwidth]{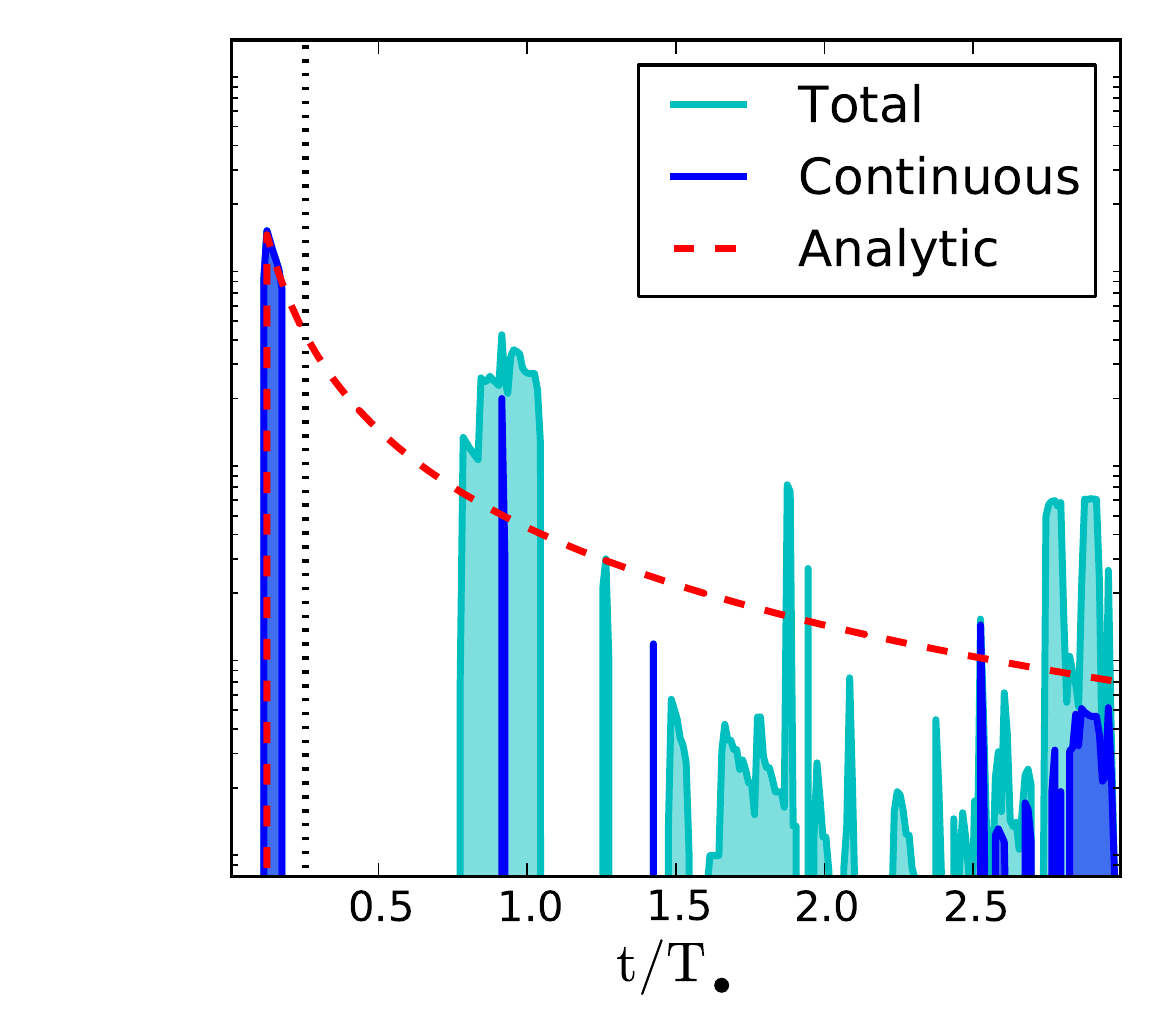} \\
\end{array}$
\caption{Continuous vs. Delayed Accretion.  In these simulations, $a=0.1$ milliparsec, $\phi=0^\circ$, $\theta=0^\circ$, and $M_\bullet=10^6 \, \mathrm{M}_\odot$.  From left to right, $q=0.01$, $q=0.1$, and $q=1$.  In light blue, we plot the total accretion onto the primary black hole, while in dark blue we overlay only the continuous accretion defined by equation \ref{eqn:continuous}.  For the smallest mass ratio, all material is accreted according to the analytic prediction for a single BH, and is therefore all considered continuous.  As $q$ increases, gaps appear in the fallback rate curves, which lead to delayed accretion.  In typical $q=0.1$ runs as in the central panel, it is common to see continuous accretion occur periodically and in agreement with the analytic power law, while delayed accretion contributes to extra bursts of accretion at random times.  For $q=1$, nearly all accretion past the first interruption of accretion is delayed, and the $t^{-5/3}$ power law is destroyed.  \label{fig:continuousTotal}}
\end{center}
\end{figure*}

In Figure \ref{fig:continuousTotal}, we plot the amount of continuous accretion versus the amount of total accretion for a set of simulations where $q$ is varied.  In these runs, $M_\bullet = 10^6 \, \mathrm{M}_\odot$, $a = 0.1$ milliparsec, $\phi = 0^\circ$, and $\theta=0^\circ$.  From left to right, $q = 0.01$, $q=0.1$, and $q=1$.  Total accretion is plotted in light blue, while continuous accretion alone is plotted in dark blue.  Accretion onto the secondary BH is not plotted here, but it is only nonzero for the $q=1$ case.  For the smallest mass ratio, accretion is never interrupted within the 3 periods simulated, in defiance of equation \ref{eqn:T_tr}.  Instead, accretion proceeds as it would around a single BH, and all material is categorised to accrete continuously.  This is often, but not always, the case in our $q=0.01$ simulations.  For larger $q$, interrupted accretion enables delayed accretion, which appears randomly in both other panels.  In typical $q=0.1$ cases such as that shown in the central panel, it is common for continuous accretion to occur on regular intervals.  In the $q=1$ case, almost all accretion following the first interruption is delayed, and the fallback rate loses resemblance with the $t^{-5/3}$ power law.  

It is interesting to notice that continuous accretion can (i) persist in the $q=0.01$ case in spite of equation \ref{eqn:T_tr} and (ii) occur regularly even if accretion is interrupted.  This can be visually confirmed by the presence of a coherent stream even late in the simulations.  Since continuously accreted material interacts minimally with the system, we predict that this accretion would be modified the least by the addition of hydrodynamics.  Delayed accretion, on the other hand, would likely be involved in additional stream intersections that are not captured in our simulations.  Because these extra intersections dissipate energy, hydrodynamics can only increase the amount of accretion onto the BHs.  We therefore conclude that delayed accretion may occur earlier and with a greater amount of material if hydrodynamical effects are taken into account, and caution readers against over-interpreting the timing and amplitude of delayed accretion.  In addition, because each gap in the fallback rate leads to more material orbiting throughout the system, fallback rate curves with larger gaps are more likely to be modified by the addition of hydrodynamics.

\subsection{Disruptions in the Plane}
\label{ssec:coplanar}

When we set $\theta = 90^\circ$, the angular momenta of the star and the secondary BH align, and all debris is confined to the plane of the MBHB.  It is in this configuration that the influence of the secondary BH is maximised, and we notice qualitative differences.  First, the secondary BH has strong close encounters with the stream, leaving distinct signatures on the resultant fallback rate.  Second, tight equal-mass binaries tend to accrete equally, and can experience what we term ``stream trading.''

\subsubsection{Close Encounters with the Secondary BH}

As explored in \S\ref{ssec:continuousDelayed}, an infalling stream persists even at late times during our simulations.  Therefore, in the plane, close encounters between the secondary BH and the stream are inevitable.  With knowledge of the initial conditions, it is straightforward to calculate the times, $t_{\mathrm{enc},j}$, that these close encounters should occur.

\begin{equation}
t_{\mathrm{enc},j} \approx T_\bullet \cdot \left[ \frac{(\phi + \phi_\mathrm{GR} + 180^\circ)\mod 360^\circ}{360^\circ} +j \right] \label{eqn:t_dir}
\end{equation}

\noindent where $j$ is a whole number.  The expression in parentheses is an approximation of the direction that the stream is launched.  $\phi$ is as usual the azimuthal coordinate of the debris at $t=0$, $180^\circ$ is added to account for the fact that the infalling stream forms at the opposite side of the disruption, and $\phi_\mathrm{GR}$ is the extra precession at pericentre due to GR.  $\phi_\mathrm{GR}$ can be written 

\begin{align}
\phi_\mathrm{GR} &= \frac{6 \pi G M_\bullet}{c^2 a_\mathrm{mb} (1 - e_\mathrm{mb}^2)} \cdot \frac{180^\circ}{\pi}\\
&= \frac{3 \pi G M_\bullet}{c^2 R_*}\left( \frac{M_*}{M_\bullet} \right)^{1/3}  \cdot \frac{180^\circ}{\pi}\label{eqn:phi_gr}
\end{align}

\noindent where $a_\mathrm{mb}$ is the semimajor axis of the most bound material and $e_\mathrm{mb}$ is its eccentricity. 

\begin{figure*}
\centering
\includegraphics[width=\textwidth]{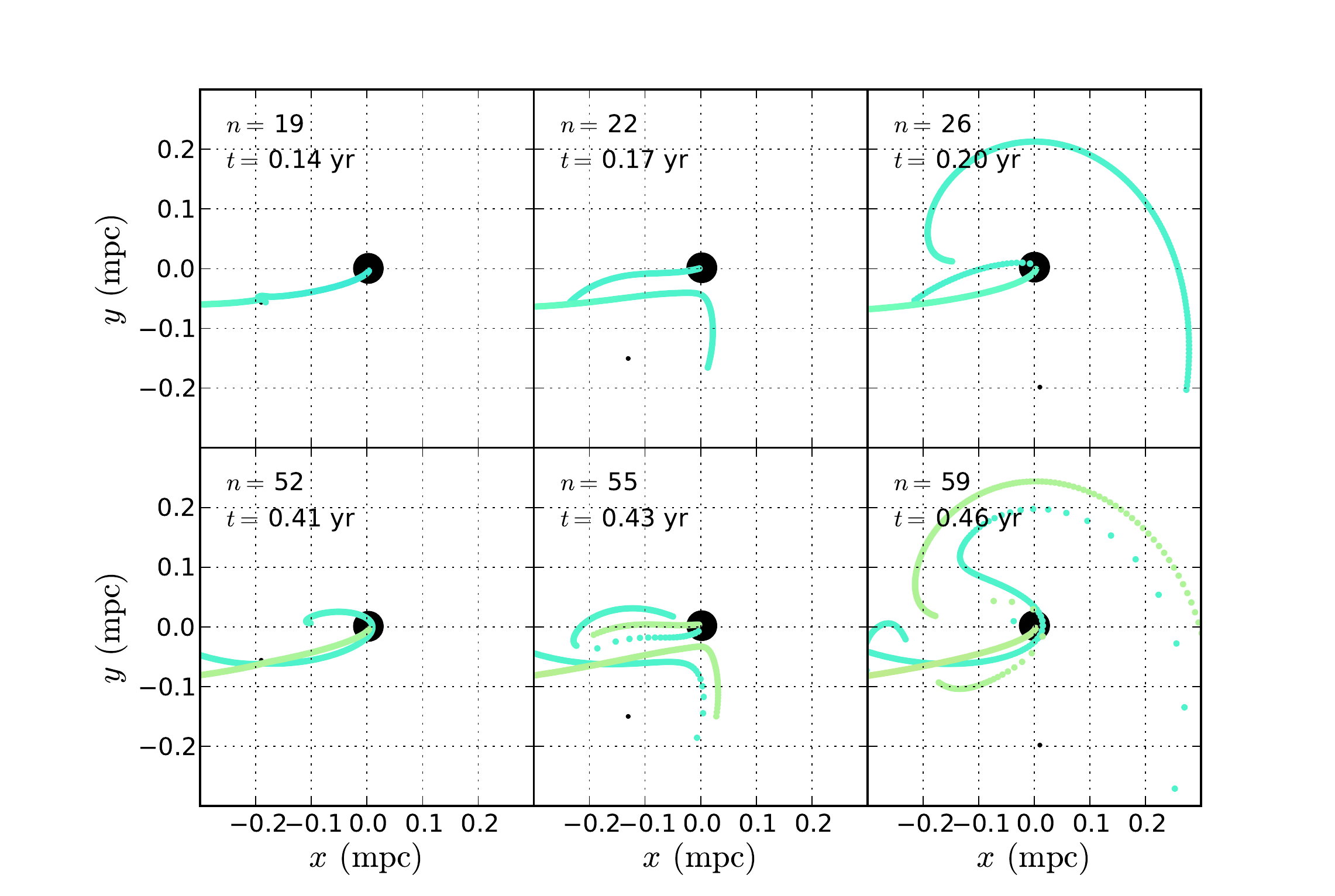}
\caption{Visualisation of direct influence of the secondary black hole.  In this run, $q=0.01$, $a=0.1$ milliparsec, $\phi=0^\circ$, $\theta=90^\circ$, and $M_\bullet=10^6 \, \mathrm{M}_\odot$.  Each time the black hole intercepts the stream, surrounding material is thrown into an expanding arc.  \label{fig:vis_direct}}
\end{figure*}

The effect of these close encounters is visualised in Figure \ref{fig:vis_direct}, where we show snapshots of the run where $q=0.01$, $a=0.1$ milliparsec, $\phi=0^\circ$, $\theta=90^\circ$, and $M_\bullet=10^6 \, \mathrm{M}_\odot$.  The secondary BH passes through the stream in the first panel.  Shown in the following two panels, material it encounters is flung into an expanding arc.  The second close encounter is illustrated in the lower panels, where there are now {\it two} streams, one formed of disturbed material that failed to accrete after the first direct encounter.  This occurs each time the secondary completes one orbit.  Despite the comparatively small mass of the secondary black hole, it has a noticeable effect on the dynamics of the stream.

\begin{figure*}
\centering
\includegraphics[width=0.5\textwidth]{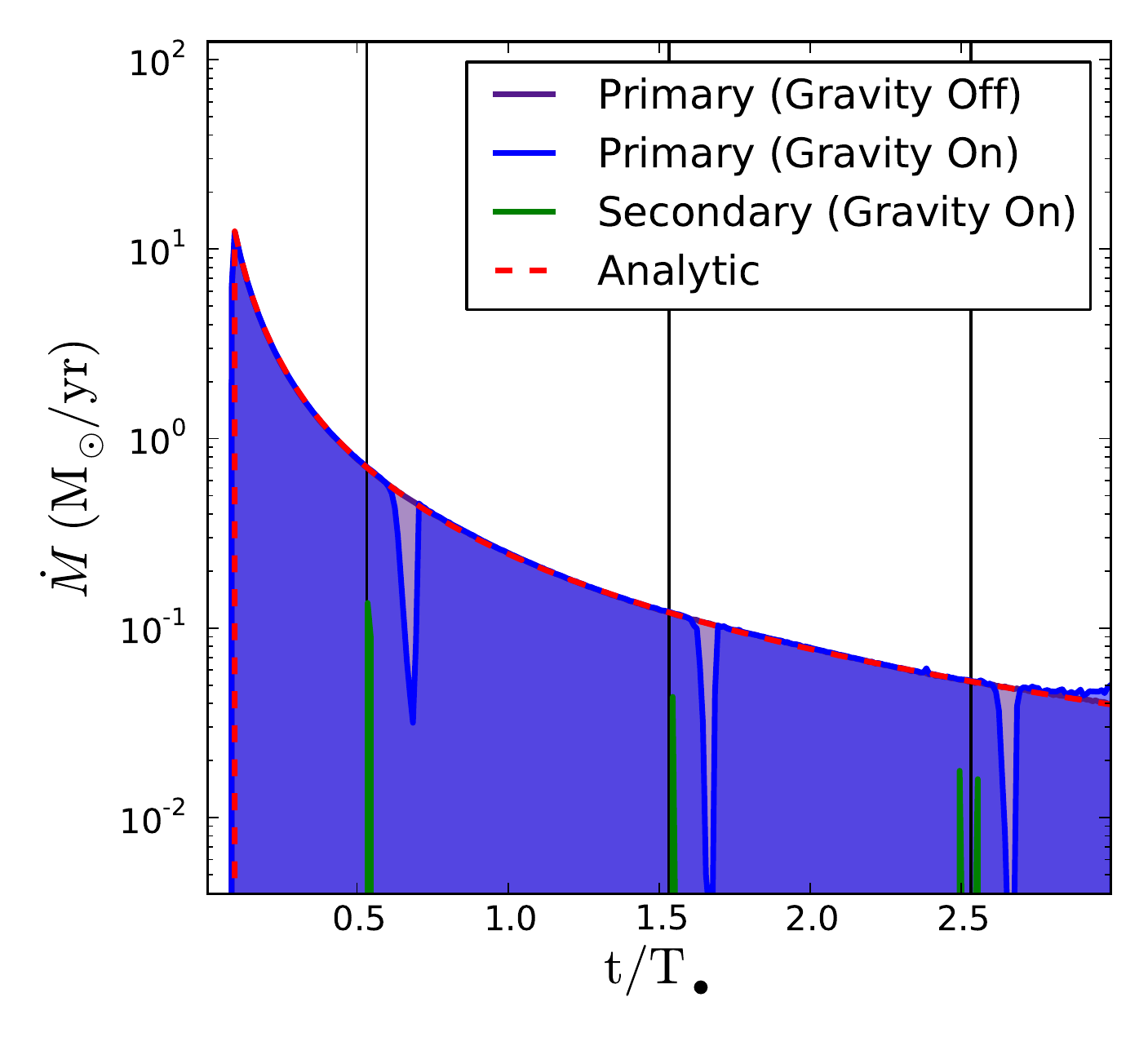}
\caption{Close encounters with the secondary black hole.  In these runs, $q=0.01$, $a=0.1$ milliparsec, $\phi=0^\circ$, $\theta=0^\circ$, and $M_\bullet=10^6 \, \mathrm{M}_\odot$.  When the secondary does not gravitationally interact with the particles, fallback occurs exactly as predicted analytically about a single black hole.  When it is turned on, periodic dips occur that are preceded by small bursts of secondary accretion.  These dips contribute to the amount of chaotic material in the simulation, which results in a small amount of extra accretion at late times. \label{fig:directInfluence}}
\end{figure*}

These close encounters can be seen in the fallback rates of these simulations.  They manifest as dips preceded by sharp bursts of accretion onto the secondary.  This effect is isolated in Figure \ref{fig:directInfluence}, where the mass accretion onto both holes is plotted.  The blue curve, which as usual displays the accretion onto the primary BH, is punctuated by sharp dips that are preceded by brief periods of secondary accretion, shown in green.  These bursts of secondary accretion are due to particles captured when the secondary BH ploughed through stream.  Solid black lines show the results of equation \ref{eqn:phi_gr}, which align nicely with these peaks of secondary accretion.  As a sanity check, we generate the fallback rate obtained in a run where the secondary BH does not interact with the particles gravitationally, but the primary BH continues to move in the binary orbit.  The result, overlaid in purple, does not show these dips, and instead agrees with the analytic curve for a single black hole.  

We expect that if hydrodynamics is taken into account, accretion onto the secondary BH will be increased and spread out in time.  Recall our simple accretion prescription:  we accrete particles onto the secondary BH if they enter within twice {\it its} tidal disruption radius, motivated only by consistency with the accretion radius of the primary BH.  During these close encounters, we observe that particles on opposite side of the secondary BH pass through each other when the BH ploughs through the stream.  A full hydrodynamic treatment is required for a more realistic estimate of the accretion onto the secondary BH during such episodes.

\subsubsection{Stream Trading with a Tight, Equal-mass Binary}

If $q \rightarrow 1$ and $a \lesssim a_\mathrm{mb}$, where $a_\mathrm{mb}$ refers to the semimajor axis of the most bound material, the primary and secondary BHs become indistinguishable.  We discover that as such a MBHB revolves, the two BHs alternate their accretion by physically exchanging the stream.  This ``stream-trading'' can be thought of as the tight, equal-mass limit of the close-encounters described in the previous section.

\begin{figure*}
\centering
\includegraphics[width=\textwidth]{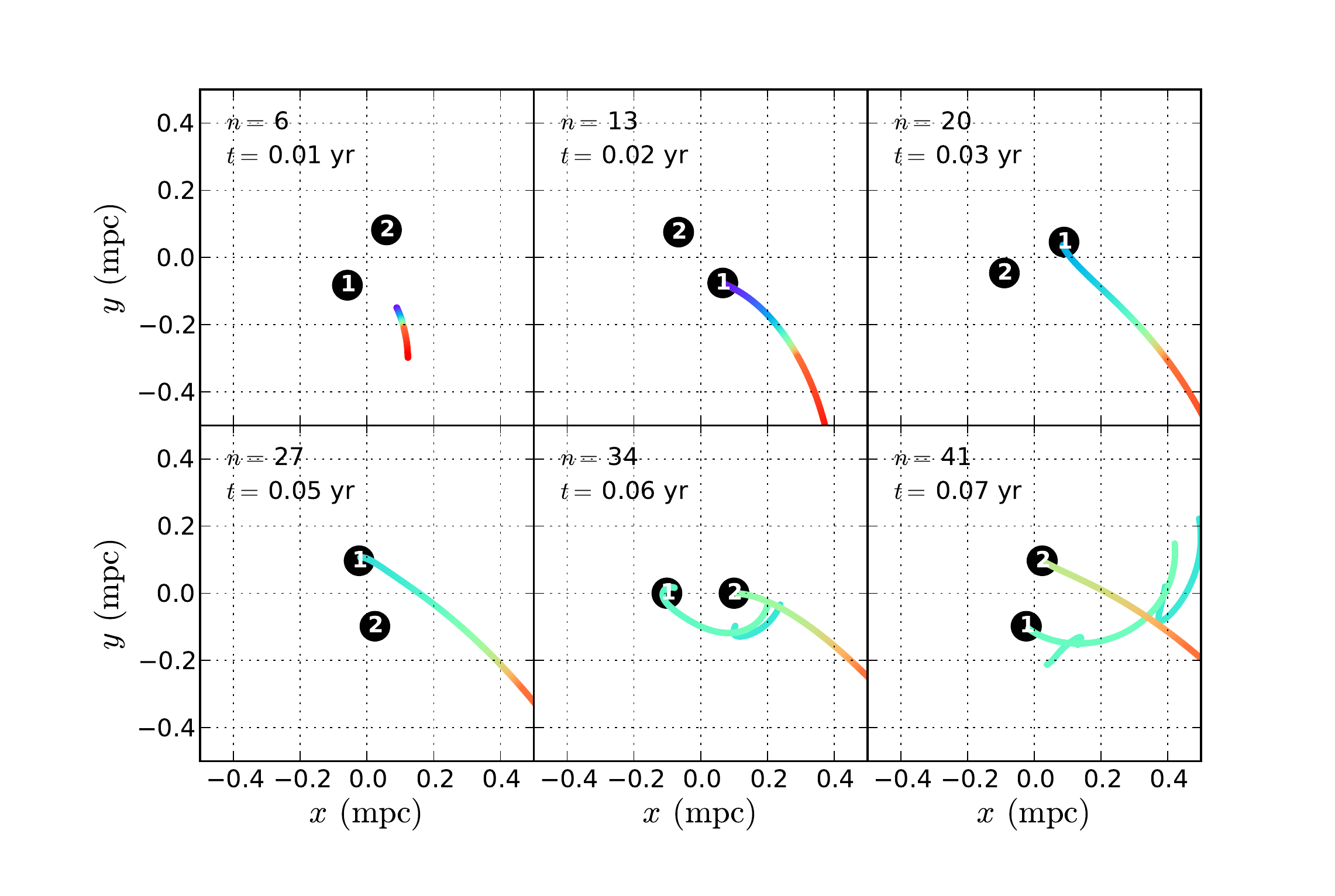}
\caption{Stream-trading with a compact equal-mass binary.  Here, $a = 0.1$ milliparsec, $\phi=120^\circ$, and $M_\bullet = 10^7 \, \mathrm{M}_\odot$.  BHs are numbered for clarity.  The stream extends and begins falling back onto the primary BH between the first and second panels.  The black holes trade ownership of the infalling stream twice a period, with the first switch occurring between the fourth and fifth panels.  The stream is traded each time the non-dominant black hole intercepts the stream. \label{fig:vis_stream_trading}}
\end{figure*}

In Figure \ref{fig:vis_stream_trading}, we show snapshots of the simulation where $a = 0.1$ milliparsec, $\phi=120^\circ$, $\theta=90^\circ$, and $M_\bullet = 10^7 \, \mathrm{M}_\odot$.  Here, we number the primary and secondary BHs for clarity, where by primary we refer to the BH from which the stream was launched.  In the first and second panels, the stream extends and begins accreting onto the primary BH.  Between the fourth and fifth panels, the secondary BH steals the infalling stream, becoming the dominant accretor.  Each time the non-dominant accretor passes through the stream, it retrieves control.  This occurs twice each period.

\begin{figure*}
\begin{center}$
\begin{array}{cc}
\includegraphics[width=0.5\textwidth]{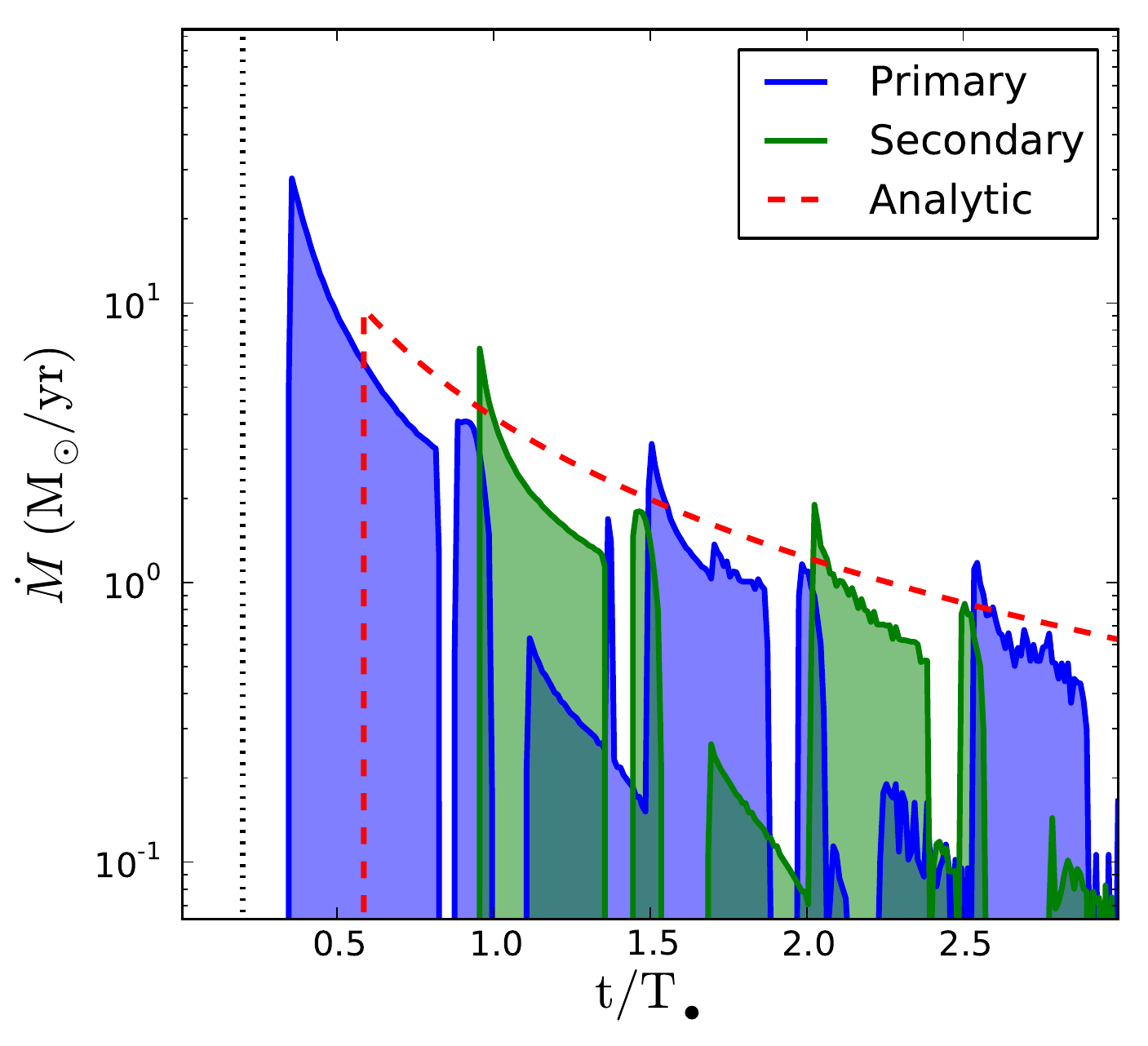} &
\hspace{-15pt}
\includegraphics[width=0.5\textwidth]{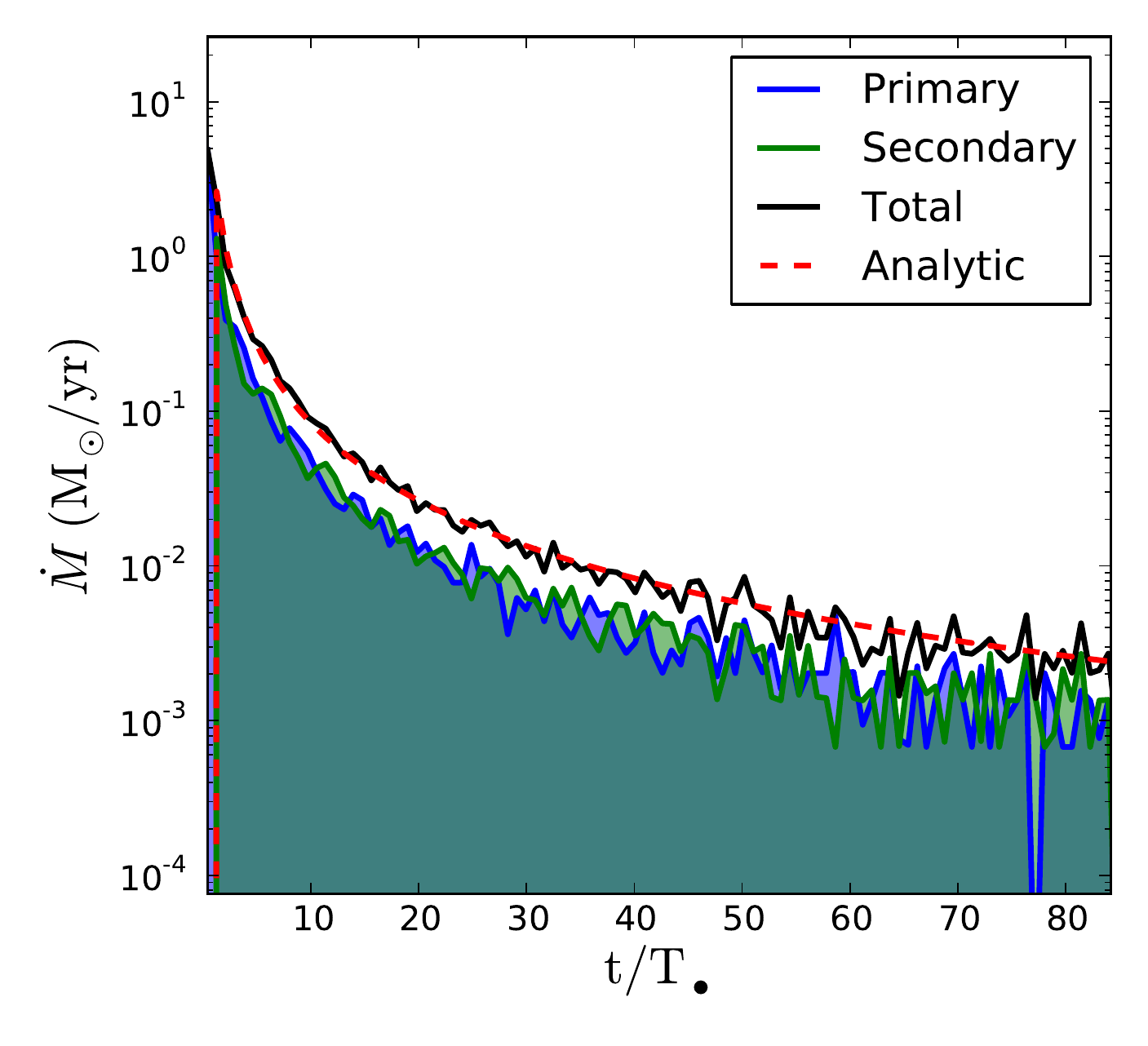} \\
\end{array}$
\caption{Mass fallback rate for a compact equal-mass binary.  Parameters are as in Figure \ref{fig:vis_stream_trading}.  On short time scales, the fallback rate is scalloped as the holes trade ownership of the stream twice a period.  On longer time scales, we can see that both holes accrete at comparable rates.   \label{fig:stream_trading}}
\end{center}
\end{figure*}

Stream trading leaves a clear signature on the mass fallback rates onto the two black holes.  This is shown over two different time scales in Figure \ref{fig:stream_trading}.  On short time scales, the fallback rate is scalloped, characterised by segments that each decay faster than $t^{-5/3}$.  On the right panel, we show the fallback rate from a simulation extended to five years.  On longer time scales, the BHs steadily accrete at comparable rates.  In addition, although either BH fails to reach the analytic curve individually, the {\it total} accretion does. Short term variability is evident, due to the scalloped fallback rate on smaller time scales.  We speculate that given the extreme velocities of these BHs, $0.098c$ in this case, future time-domain data may reveal spectral variability as a result of stream-trading.  

\begin{figure*}
\centering
\includegraphics[width=\textwidth]{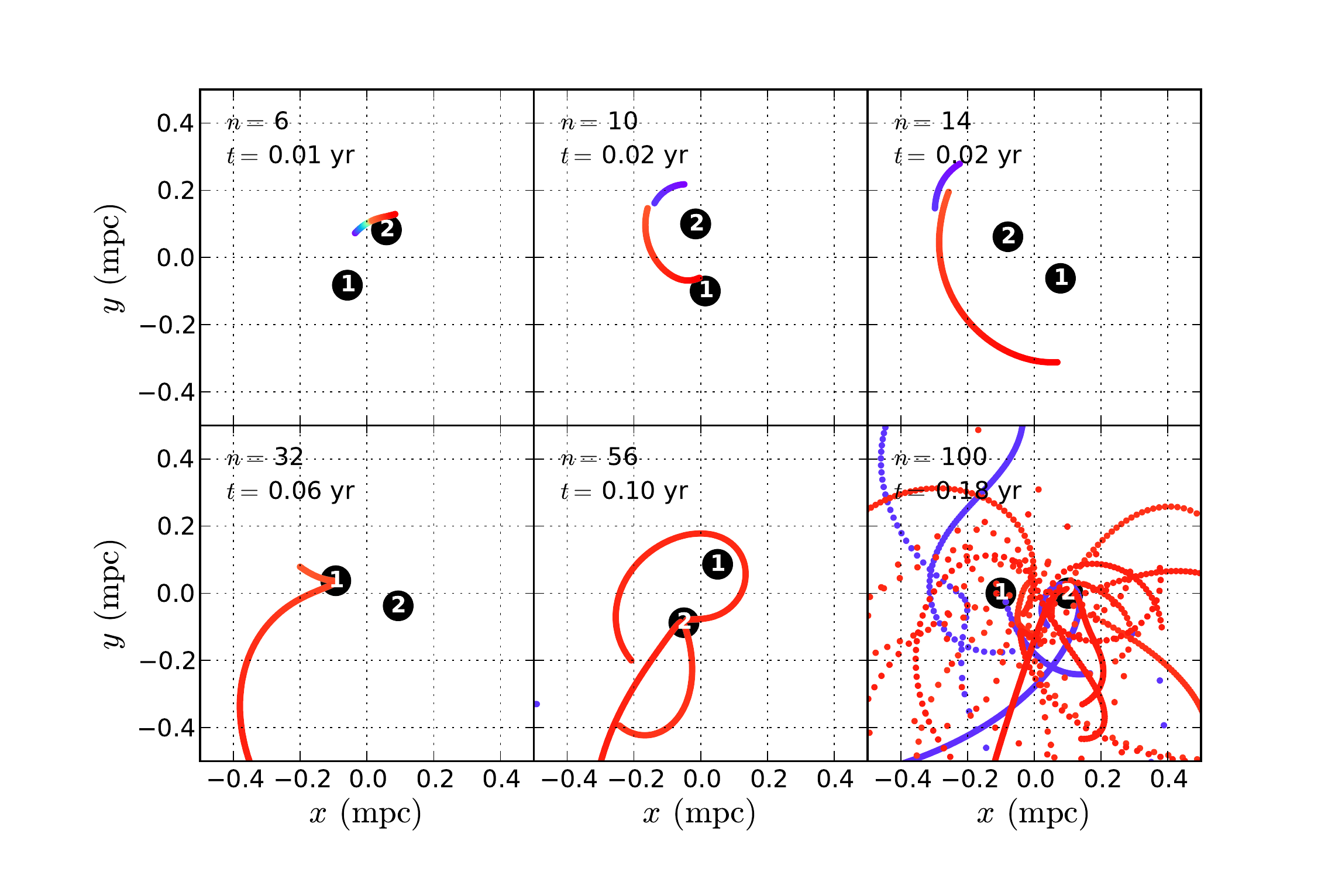}
\caption{A special case where most of the material promptly falls into the {\it secondary} black hole. Parameters are as in Figure \ref{fig:vis_stream_trading}, except with $\phi=240^\circ$.  BHs are numbered for clarity.  Between the first two panels, the secondary black hole engulfs nearly the entire stream, and the remaining material is flung to opposite sides of the hole.\label{fig:vis_special}}
\end{figure*}

\begin{figure*}
\centering
\includegraphics[width=0.5\textwidth]{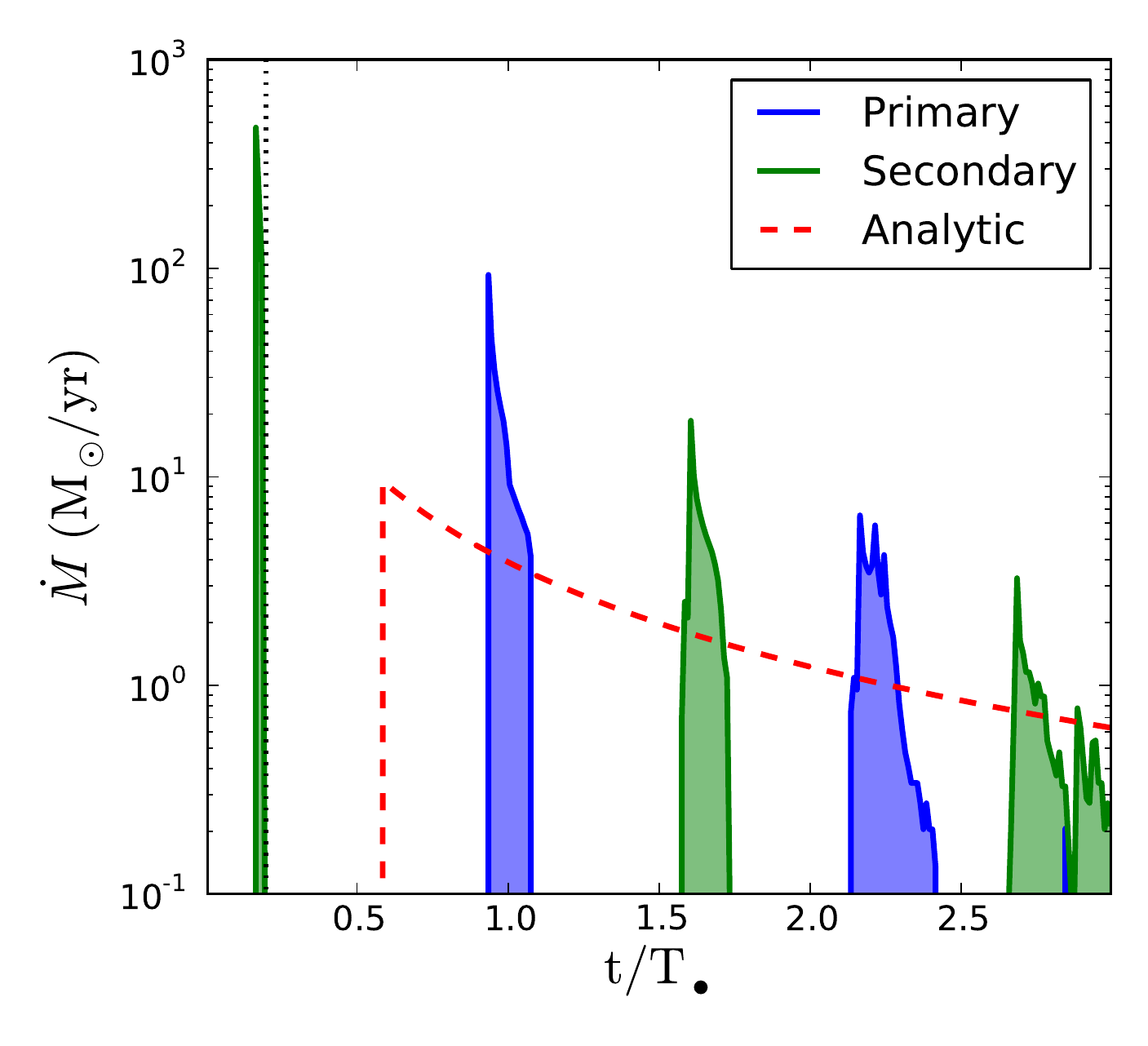}
\caption{Mass fallback rate for the special case visualised in Figure \ref{fig:vis_special}.  A large fraction of the accretion occurs in a massive spike before $t_\mathrm{min}$. \label{fig:special}}
\end{figure*}

A similar but special case is shown in Figure \ref{fig:vis_special}.  Here, we have only changed $\phi$ from $120^\circ$ to $240^\circ$.  As shown in the first panel, the secondary black hole directly interacts with the entire stream before it is able to stretch out very far.  Immediately, it consumes most of the matter, while the matter that remains is slingshot around the BH.  Note that in the second panel, the entire stream has passed through itself, swapping the positions of the bound and unbound sections of the stream.  In a hydrodynamical simulation, we would expect that this should cause a violent shock.  As the simulation continues, the remainder of the material is accreted as in the previous case, as shown by its mass fallback rate curve in Figure \ref{fig:special}.  After the 3 periods simulated, the primary BH accretes $0.21 \, \mathrm{M}_\odot$, while the secondary BH accretes $0.54 \, \mathrm{M}_\odot$.  Given that the entirety of the star collides at once in the vicinity of the secondary black hole, we doubt that our ballistic simulations capture in detail the true accretion rates onto either BH.  Nevertheless, we find it noteworthy that it is possible for the secondary BH to accrete more mass than the primary.

\section{Special Investigations}
\label{sec:special}

\subsection{Timing of Truncation}

L09 and L14 noticed discrepancies between the timing of the first interruption and the analytic predictions of equation \ref{eqn:T_tr}.  We have even noted in the previous section that it is possible for truncation not to occur at all.  We therefore investigate the timing of truncation as a function of $q$ and $\theta$, which appears explicitly in equation \ref{eqn:T_tr}, but also $\phi$, which does not.  Simulations are run for $M_\bullet = 10^7 \, \mathrm{M}_\odot$, $a=1$ milliparsec, $\phi \in [0^\circ,360^\circ)$, $\theta \in \{0^\circ, 45^\circ, 90^\circ\}$, and $q \in \{0.01,0.1,1\}$.  The $\phi$ coordinate is sampled much more finely than in our usual grid, using steps of $6^\circ$.   For each run, we define the truncation time $T_\mathrm{tr}$ as the first time bin where the mass fallback rate onto the primary black hole is equal to 0.  These simulations are run for only two periods, and with only $10^4$ particles each, since we are interested neither in resolving fallback rate curves nor in behaviour at late times.

\begin{figure*}
\centering
\includegraphics[width=\textwidth]{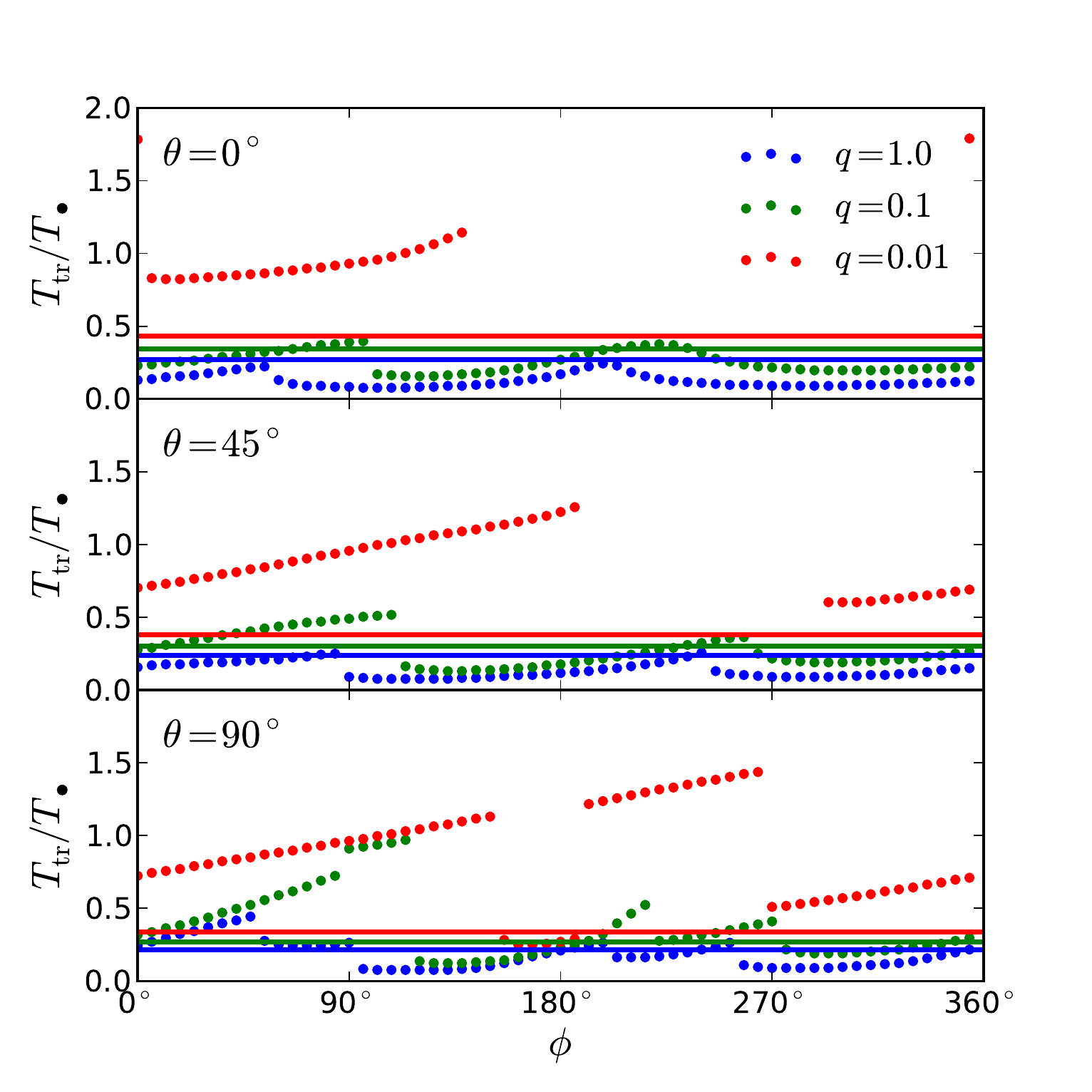}
\caption{Timing of first interruption of accretion as a function of $\phi$ for select $q$ and $\theta$.   $M_\bullet = 10^7 \, \mathrm{M}_\odot$ and $a=1$ milliparsec in all simulations.  Circles represent the time of first interruption as a function of $\phi$, while solid lines show the estimates for $T_\mathrm{tr}$ in equation \ref{eqn:T_tr}.  Evidently, equation \ref{eqn:T_tr} performs poorly, especially at $q=0.01$.  There is also clear $\phi$-dependence, with discontinuities that do not appear in a regular manner.  More discontinuities are present in the plane, when $\theta=90^\circ$.  For $q=0.01$, there are some values of $\phi$ which do not experience interruptions at all within 2 periods.  \label{fig:truncationFinder}}
\end{figure*}

Results are shown in Figure \ref{fig:truncationFinder}.  Each circle plots $T_\mathrm{tr}/T_\bullet$ for each run, while solid lines mark analytic predictions of $T_\mathrm{tr}$ by equation \ref{eqn:T_tr}.  Blue, green, and red points represent runs with $q$ equal to 1.0, 0.1, and 0.01 respectively.  From top to bottom, these runs correspond to $\theta$ equal to 0, 45, and 90 degrees respectively.  We notice significant $\phi$-dependence, despite its absence in equation \ref{eqn:T_tr}.  $T_\mathrm{tr}$ appears to have several discontinuities as a function of $\phi$, although we notice no clear pattern.  There are more discontinuities at $\phi=90^\circ$, likely due to close encounters with the secondary BH.  Equation \ref{eqn:T_tr} performs most poorly for our extreme mass ratio, $q=0.01$.  For several of the runs where $q=0.01$ and $\theta \neq 90^\circ$, truncation never occurs at all.  In the runs where it does, $q=0.01$ values differ from the analytic estimate by an average of $0.56 T_\bullet$ when $\theta=0^\circ$, $0.50 T_\bullet$ when $\theta=45^\circ$, and $0.53 T_\bullet$ when $\theta=90^\circ$.  In addition, the equation tends to overestimate $T_\mathrm{tr}$ for $q=1$.  This is worst at $\theta=0^\circ$, where $T_\mathrm{tr}$ for $q=0$ is overestimated by an average of $0.15 T_\bullet$.

We have determined that the timing of truncation depends on $\phi$ in a complicated manner.  The analytic estimate for $T_\mathrm{tr}$ via equation \ref{eqn:T_tr} fails for extreme mass ratios and does not account for this $\phi$-dependence.  Note that these results are dependent on the accretion radius, which we study in \S\ref{ssec:r_acc}.

\subsection{Sensitivity to $r_\mathrm{acc}$}
\label{ssec:r_acc}

One of our key assumptions in the immediate accretion of any material that enters within $r_\mathrm{acc}$ of either BH, which is set to twice the tidal disruption radius of either hole.  To test the sensitivity of our results to this choice of $r_\mathrm{acc}$, we perform an experiment in which the same simulation is run three times, where $r_\mathrm{acc}$ is decreased from 4 to 2 to 1 times the tidal disruption radius.  (In practice, this last value is set to 1.01 to allow particles to both exit and re-enter.)  In this experiment, we choose the parameters $M_\bullet = 10^6 \, \mathrm{M}_\odot$, $a = 1$ milliparsec, $q=0.1$, $\phi=0^\circ$, and $\theta=0^\circ$.  

\begin{figure*}
\begin{center}$
\begin{array}{ccc}
\includegraphics[width=0.35\textwidth]{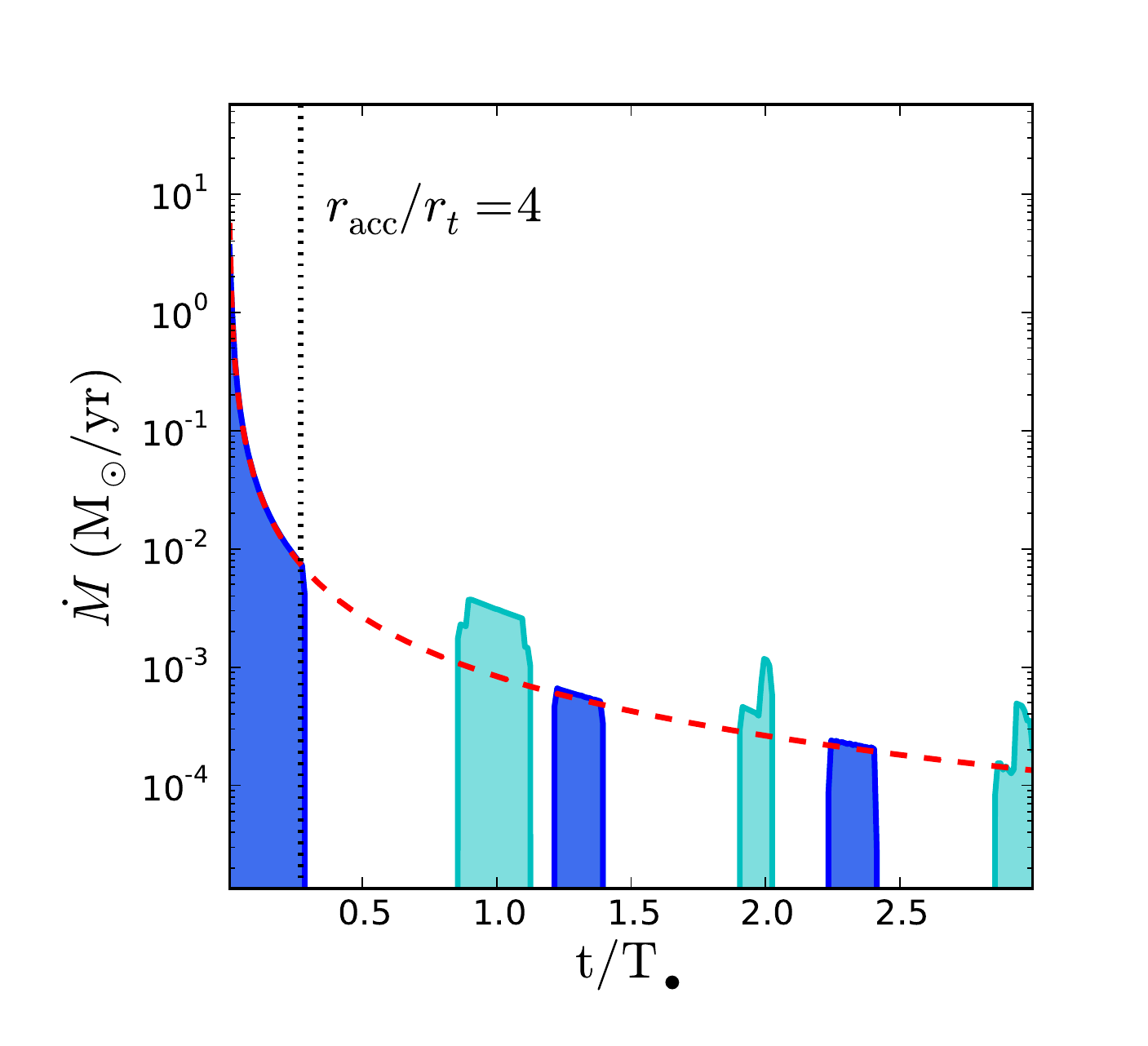} &
\hspace{-26pt} 
\includegraphics[width=0.35\textwidth]{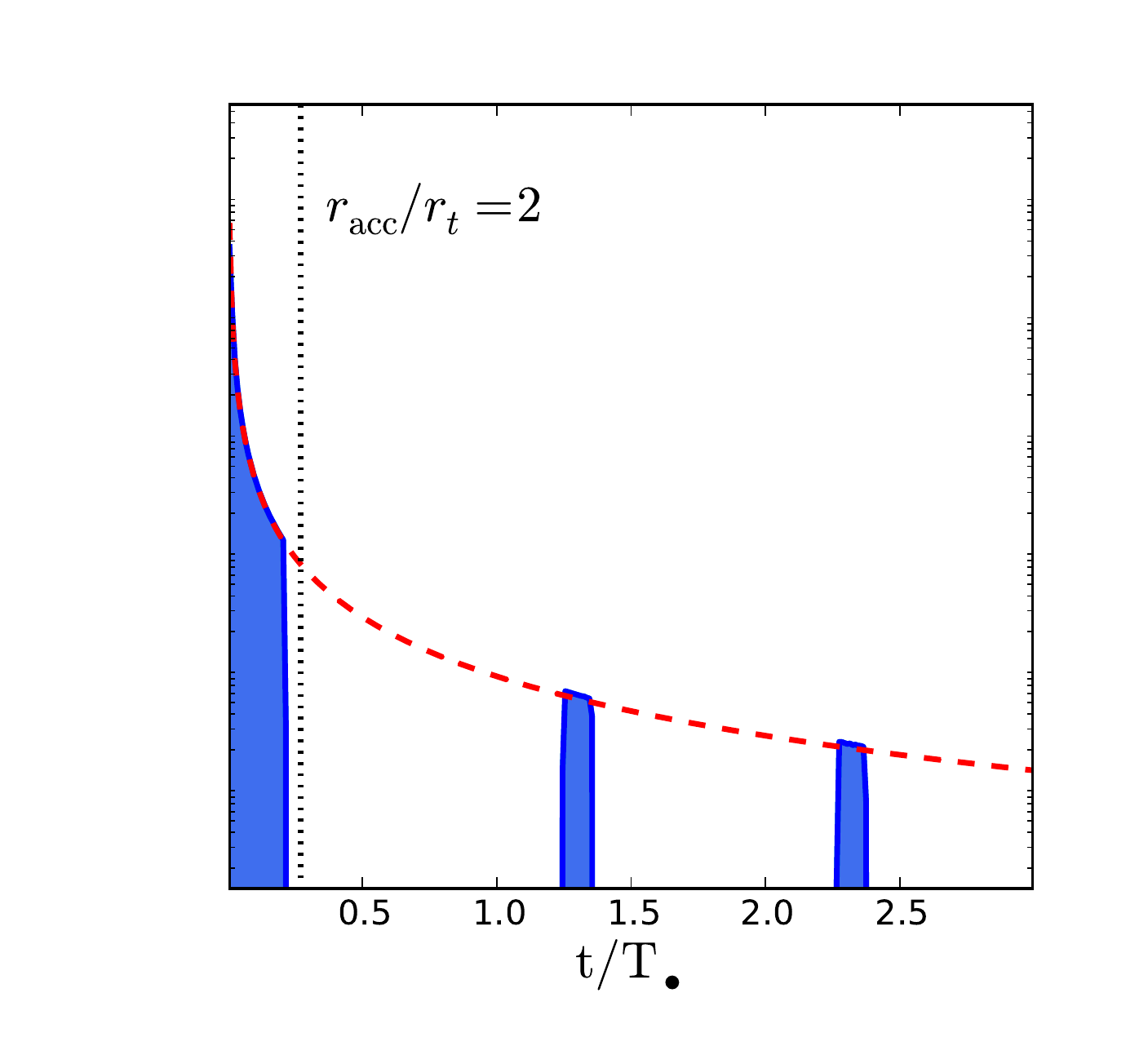} &
\hspace{-26pt}
\includegraphics[width=0.35\textwidth]{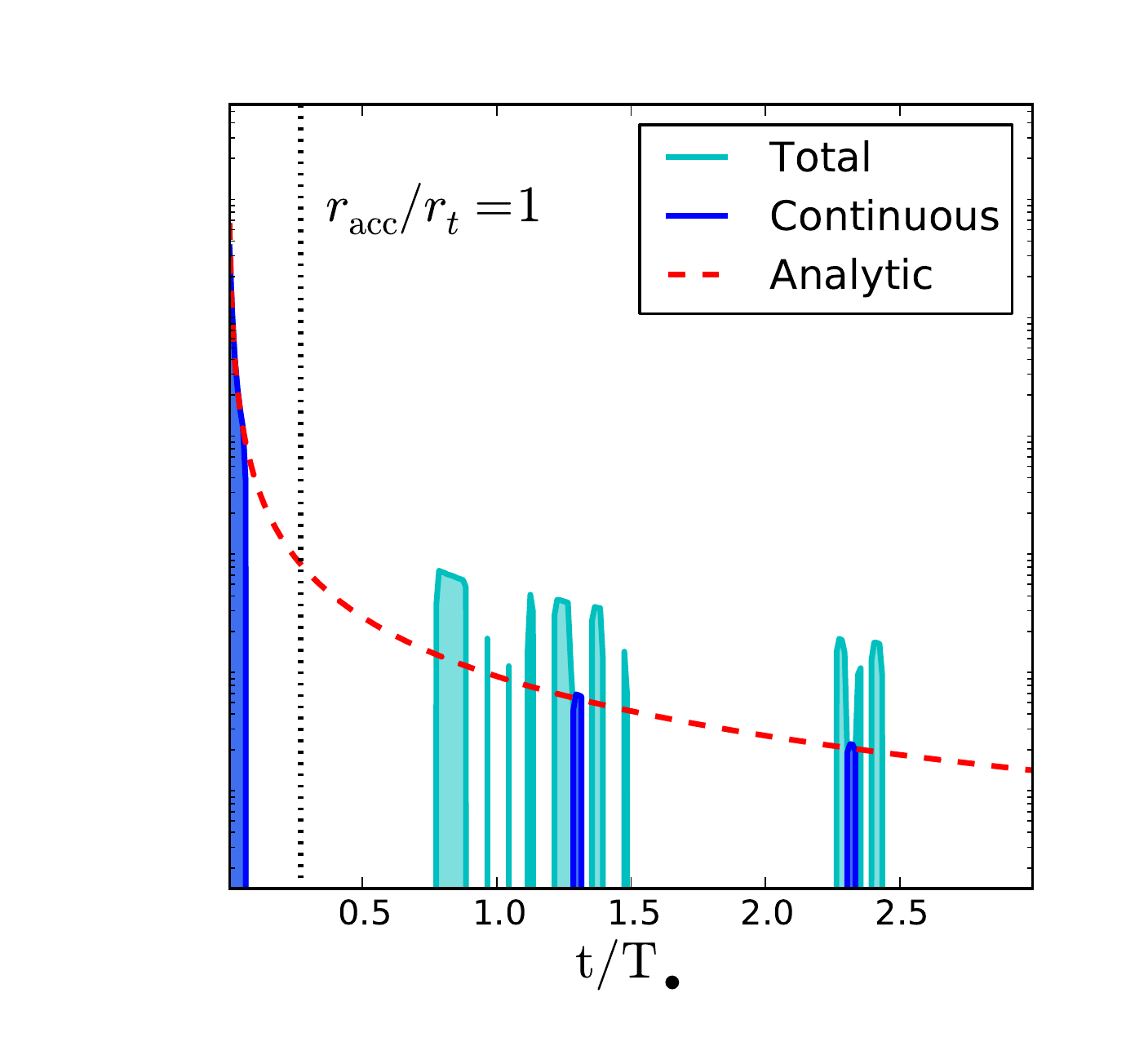} \\
\end{array}$
\caption{Sensitivity to $r_\mathrm{acc}$.  In this experiment, the same simulation is run with different values of the accretion radius, $r_\mathrm{acc}$.  From left to right, $r_\mathrm{acc}$ is set to 4, 2, and 1 times $r_t$ respectively.  As in Figure \ref{fig:continuousTotal}, the light blue curve plots total accretion, while the darker blue curve plots only continuous accretion.  As $r_t$ is decreased, the periods of continuous accretion shrink.  No obvious pattern occurs with delayed accretion.  \label{fig:testAccretionRadius}}
\end{center}
\end{figure*}

Results are shown in Figure \ref{fig:testAccretionRadius}.  As in Figure \ref{fig:continuousTotal}, we plot total accretion in light blue and continuous accretion only in dark blue.  From left to right, $r_\mathrm{acc}$ is set to 4, 2, and 1 times $r_t$ respectively.  We find that changing the value of $r_\mathrm{acc}$ can have drastic effects on the fallback rate.  In fact, the similarity between the three cases is not apparent without dividing the accretion into continuous and delayed components.  Periods of continuous accretion appear to occur at the same times, yet their duration shrinks with decreasing $r_\mathrm{acc}$.  There is no obvious pattern to delayed accretion due to two competing effects.  Decreasing $r_\mathrm{acc}$ makes it more difficult for both forms of accretion to occur.  However, decreasing $r_\mathrm{acc}$ also widens the temporal gaps in the fallback rate, which increases the amount of material that could possibly contribute to delayed accretion.

Ideally, $r_\mathrm{acc}$ relates to where, and how fast, circularisation occurs.  If circularisation occurs very close to pericentre, as may be facilitated by nozzle shocks \citep{Guillochon+2014} or close stream intersections when apsidal precession is strong \citep{Dai+2015}, then $r_\mathrm{acc} \sim r_t$.  Otherwise, intersections can occur far from the BH \citep{Shiokawa+2015} and $r_\mathrm{acc} \gg r_t$.  In addition, circularisation would be slow in this case and the light curve may not follow the fallback curve in Figure \ref{fig:testAccretionRadius}.  As the fallback rate is very sensitive to $r_\mathrm{acc}$, future hydrodynamical simulations will enlighten us on the immediate peak accretion and long-term behaviour of debris dynamics in these MBHB TDEs.

In summary, decreasing $r_\mathrm{acc}$ decreases periods of continuous accretion predictably, but adds or subtracts delayed accretion randomly.  We conclude that fallback rates generated this way are subject to substantial uncertainties due to the assumption of a fixed accretion radius.  Even the time of first interruption depends on the value chosen for $r_\mathrm{acc}$.  We therefore implore caution when applying such fallback rates that are generated without hydrodynamics.

\subsection{Where do stream crossings occur?}

As mentioned earlier in this paper, stream self-intersections are important for circularisation.  Indeed, throughout this work, we have assumed that such crossings are efficient at dissipating energy within the accretion radius of either black hole.  In this section, we relax the assumption that material is accreted within $r_\mathrm{acc}$.  Instead, we explore where stream crossings occur and discuss the implications for circularisation.

We perform two special runs where stream self-intersections are located for each snapshot.  We choose typical parameters, $M_\bullet = 10^6 \, \mathrm{M}_\odot$, $a = 1$ milliparsec, $q=0.1$, $\phi=120^\circ$, and $\theta=90^\circ$.  Since we set $\theta=90^\circ$, material is confined to the plane and thus stream intersections are well-defined without having to specify a stream width.  In these runs, 200 snapshots are saved instead of the usual 100.  We also use a special particle removal scheme in order to isolate the intersections which would occur first in a particle's history, which are most relevant for future hydrodynamical studies.  Particles are removed from the simulation only if either (i) they pass within the innermost bound circular orbit (IBCO) of a BH ($r_\mathrm{IBCO} = 4GM_\bullet/c^2$ for a non-rotating BH), or (ii) they complete 1.5 periods, calculated as we do for continuous accretion via equation \ref{eqn:continuous}.  The first condition removes particles that cause the time step to shrink infinitely small, while the second condition attempts to remove particles that probably should have already accreted, which would cause us to identify spurious intersections if left in the simulation.  In the first of these special runs, the secondary BH does not interact with the particles gravitationally.  This allows us to identify differences in behaviour caused purely by the motion of the primary BH.  In the second run, the secondary BH's gravity is switched on again as usual.

\begin{figure*}
\begin{center}
\includegraphics[width=\textwidth]{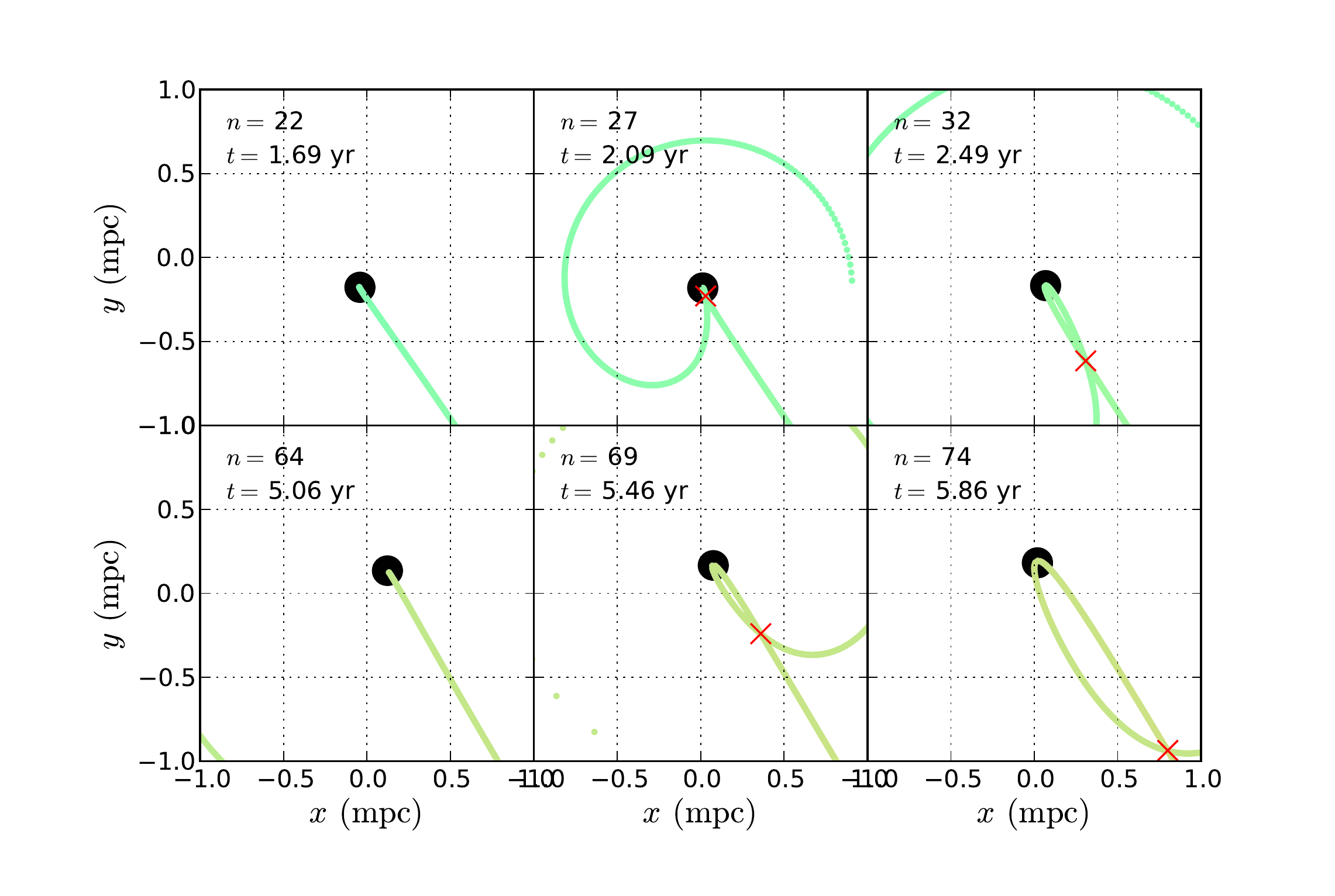}
\caption{Visualisation of our intersection experiment with the secondary BH's gravity switched off.  Intersections are marked with red ``x''s.  As the primary BH moves, the distance of closest approach varies.  Hence, apsidal precession varies in strength, and the location of first intersection changes as a function of time.  Some material can also plunge directly into the IBCO.  Note that between the two rows, the angular momentum of the infalling material has changed sign.  \label{fig:vis_int_off}}
\end{center}
\end{figure*}

In Figure \ref{fig:vis_int_off}, we visualise the results of the run where the secondary BH does not interact with the particles gravitationally.  Here, we zoom in on the central milliparsec and mark intersections with red ``x''s.  As the BH orbits the binary centre of mass (the origin), the stream's distance of closest approach varies as a function of time.  The amount of apsidal precession experienced by infalling particles varies accordingly.  As a result, unlike the case of a single BH, the location of intersection oscillates with time.  The two rows shown here depict two of these oscillations.  In the left-most panels, particles actually plunge directly into the IBCO.  In subsequent frame, the distance of closest approach increases, and particles are flung around the BH instead of being immediately accreted.  Between these two rows, although it is not always the case, the angular momentum of the infalling particles has also switched sign.  

\begin{figure*}
\centering
\includegraphics[width=\textwidth]{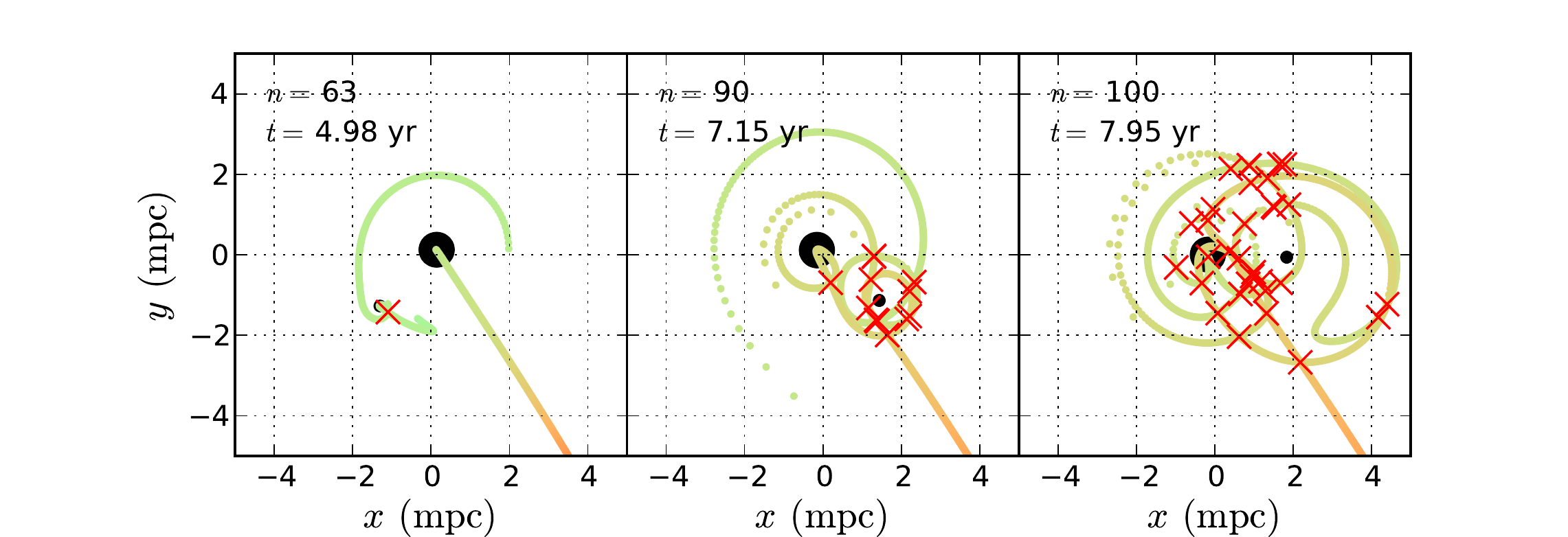}
\caption{Visualisation of our intersection experiment with the secondary BH's gravity switched on.  Material is flung around the secondary BH during many close encounters, resulting in a myriad of additional intersections.  \label{fig:vis_int_on}}
\end{figure*}

When the secondary BH's gravity is turned on, close encounters between the stream and the secondary BH create a myriad of additional intersections, visualised in Figure \ref{fig:vis_int_on}.  These begin once the apocentre of particles involved in intersection approaches 2 milliparsec, the distance between the two BHs.  In the first panel, the secondary BH interacts with a section of stream which is substantially diverted from its initial path by apsidal precession.  In the second and third panels, the BH ploughs through the infalling stream, resulting in several layers of circularly displaced debris.  As we have seen in \S\ref{ssec:coplanar}, these arcs are formed as particles are flung around the secondary BH.

\begin{figure*}
\begin{center}$
\begin{array}{ccc}
\includegraphics[width=0.5\textwidth]{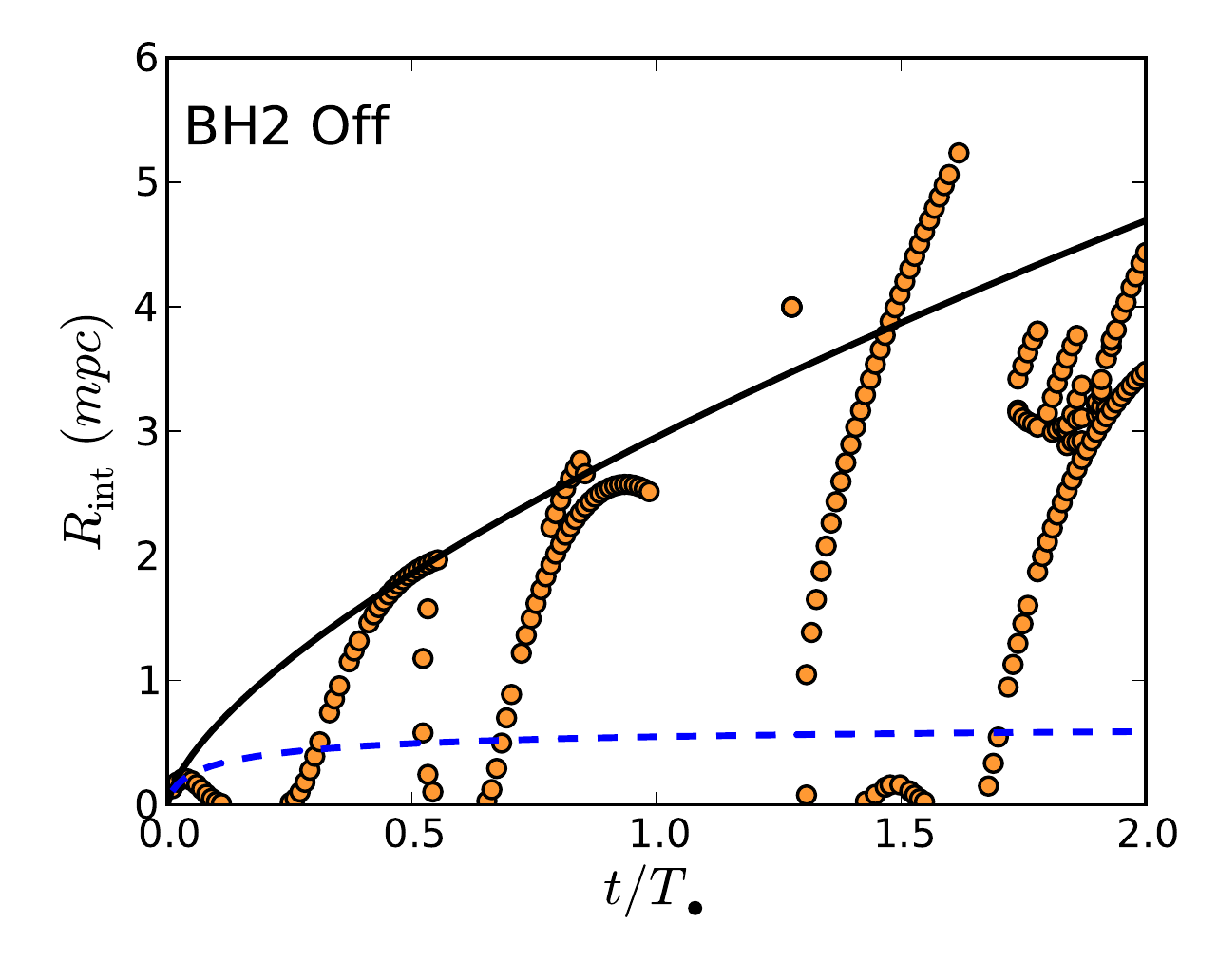} &
\hspace{-20pt} 
\includegraphics[width=0.5\textwidth]{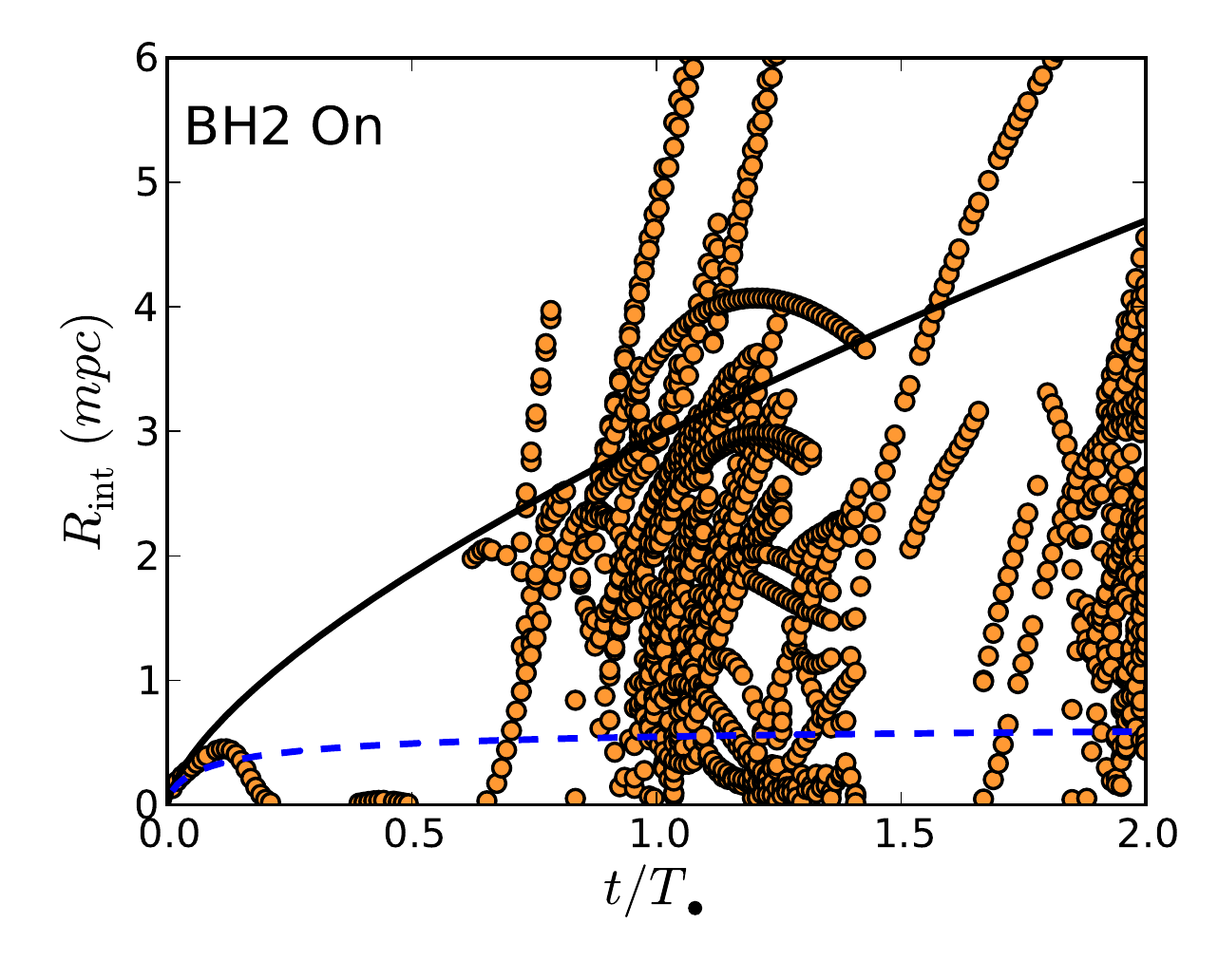} \\
\end{array}$
\caption{Intersection distances as a function of time.  Here, $M_\bullet = 10^6 \, \mathrm{M}_\odot$, $a = 1$ milliparsec, $q=0.1$, $\phi=120^\circ$, and $\theta=90^\circ$.  The secondary BH does not interact with the particles in the left panel, and its interactions are turned back on in the right panel.  The blue dashed curve corresponds to the distance at which intersections would occur in the absence of the secondary BH, while the black curve represents the curve that would be followed if all intersections occurred at apocentre around a single BH.  Substantial changes in the intersection distance occur as a result of the motion of the primary BH.  When the secondary's gravity is switched on, many more intersections occur as material is flung around it.  \label{fig:intersections}}
\end{center}
\end{figure*}

In Figure \ref{fig:intersections}, we plot the distance away from the primary BH at which intersections occur as a function of time, represented by orange circles.  The left panel shows the results of the run where the secondary BH does not interact with the particles, while the right panel shows the results when its gravity is switched back on.  Two analytic curves are also plotted.  The solid black curve represents the apocentric distance of a particle with a period equal to 2/3 the current simulation time; if all intersections occurred at apocentre, they would trace this curve.  The dashed blue curve shows the approximate location of intersection around a single BH of mass $10^6 \, \mathrm{M}_\odot$ due to apsidal precession; in the absence of the secondary BH, all intersections would trace this curve.  This latter curve is calculated via 

\begin{equation}
R_\mathrm{int} \approx \frac{[1+e(2t/3)]r_t}{1 - e(2t/3)\cos\{[\phi_\mathrm{GR}(2t/3)]/2\}} \label{eqn:R_int}
\end{equation}

\noindent where $e(T)$ and $\phi_\mathrm{GR}(T)$ represent the eccentricity and apsidal precession of a particle that has period $T$ \citep{Dai+2015}.  Since particles would normally collide around apocentre for a BH of this mass, and reaching apocentre takes 3/2 of a period, this curve is generated with $T = 2t/3$, where $t$ is the current simulation time.  We verify that this curve is matched using separate runs without a secondary BH.

Figure \ref{fig:intersections} illustrates that the location of intersection for the case of binaries (orange circles) differs substantially from the case with a single BH (blue dashed line).  This may be unintuitive since one only expects $11^\circ$ of apsidal precession due to GR for a BH of $10^6 \, \mathrm{M}_\odot$ with a penetration parameter of $\beta=1$.  Yet the left panel reveals that the motion of the primary BH is capable of inducing large fluctuations in precession angle, and therefore intersection distance.  We repeated this exercise with a Newtonian potential, and found that some distant, weak intersections still occur as a result of the time-varying potential, but none of the nearby intersections and dramatic oscillations observed here.  

In summary, the motion of the primary BH can result in substantial changes to the location of closest approach for the infalling debris, which causes the distance of intersection to vary with time.  This can make apsidal precession important even for relatively modest $M_\bullet$ and $\beta$.  In addition, encounters with the secondary BH introduce many more potential shocks to reduce orbital energy of the stream's constituent particles.  For $\theta < 90^\circ$, we expect that this latter effect will be less pronounced.  The resulting pattern of intersections should behave intermediate to the two panels in Figure \ref{fig:intersections}.  Note that these results were obtained without the use of an accretion radius, and should not be strongly affected by our lack of hydrodynamics.

\section{Discussion}
\label{sec:discussion}

\subsection{Effects of Each Parameter}

We present a synopsis of the results of our extensive parameter study below.  Our exploration focuses on the impact of these parameters on the fallback rate.

\subsubsection{The Primary Mass, $M_\bullet$}

The mass of the primary BH, $M_\bullet$, enters our model in three fundamental ways:
\begin{itemize}
\setlength{\parskip}{0cm}
\item The period of the MBHB $T_\bullet \propto M_\bullet^{-1/2}$.  This changes the timescale over which the potential varies and the secondary BH interacts with the stream.
\item The magnitude of apsidal precession $\phi_\mathrm{GR} \propto M_\bullet^{2/3}$.  (See equation \ref{eqn:phi_gr}.)  This changes the direction that the initial stream is ejected, and the degree of deflection during subsequent orbits.
\item The minimum return time $t_\mathrm{min} \propto M_\bullet^{1/2}$ to first order.  (See equations \ref{eqn:t_min} and \ref{eqn:DeltaE_Wegg}.)  This affects how quickly the flare dims.
\end{itemize}

In short, increasing $M_\bullet$ decreases the timescale over which MBHB effects are seen, increases the amount of GR apsidal precession, and increases the timescale over which the flare dims.  With other parameters fixed, it becomes more difficult to observe binary signatures as $M_\bullet$ decreases. 

While outside of our model, $M_\bullet$ also determines whether or not the fallback rate is super-Eddington, with less massive black holes having more super-Eddington flows \citep{Evans&Kochanek1989}.  Simulations suggest that such super-Eddington disks are thick and can generate strong winds and even jets around spinning BHs \citep{Sadowski&Narayan2015,Jiang+2014,McKinney+2015}.

\subsubsection{The BH Mass Ratio, $q$}

As $q \rightarrow 1$, truncation occurs sooner and gaps in the fallback rate widen.  More material is thrown onto chaotic orbits and less accretion occurs continuously, as discussed in \S\ref{ssec:continuousDelayed}.  If the TDE occurs in the plane of the binary, $q$ also determines the strength of close encounters between the secondary BH and the stream.  In the extreme case where $q=1$ and $a$ is comparable to the semimajor axis of the most bound stellar material, the two MBHs become indistinguishable and periodically trade the stream.

MBHBs with $q \sim 0.1$ are probably the most likely to be observed.  Minor galaxy mergers occur much more frequently than major mergers \citep{Lotz+2011}, and BH mass is correlated with galaxy bulge mass \citep{Kormendy&Richstone1995,Kormendy&Ho2013,McConnell&Ma2013}.  In addition, simulations of galaxy mergers that carefully track accretion onto the SMBHs reveal that large mass ratios tend to shrink, while small mass ratios tend to grow \citep{Capelo+2015}.  The dynamical friction process that brings secondary BHs to the center of merged galaxies is thought to be ineffective for extreme mass ratios, disallowing the formation of binaries with $q \lesssim 0.1$ \citep{Callegari+2011}.  On the other hand, since the lifetime of a MBHB that decays from an initial separation $R$ via gravitational radiation is given by 

\begin{equation}
t = \frac{5}{256}\frac{c^5}{G^3}\frac{R^4}{(M_1M_2)(M_1+M_2)} \propto \frac{1}{M_\bullet^3q(1+q)} \label{eqn:gw}
\end{equation}

\noindent at fixed mass, binaries with larger $q$ have shorter lifetimes due to orbital decay.

\subsubsection{The BH Semimajor Axis, $a$}

For the effect of a secondary BH to be apparent on the dynamics of the stream, the binary must have a period comparable to the decay timescale of TDE flare.  For MBHs of $M_\bullet \sim 10^{6-7} \, \mathrm{M}_\odot$, this restricts $a$ to milliparsec separations.  If $a$ becomes comparable to the semimajor axis of the most bound stellar material, it can be disturbed by the secondary BH before the most bound material accretes.  This can result in stream-trading when $q = 1$, or if the debris is fortuitously ejected in the direction of the secondary BH, a burst of accretion by the secondary BH.

\subsubsection{The Initial Position of the Star, $\phi$ and $\theta$}

Although $\phi$ does not explicitly appear in equation \ref{eqn:T_tr}, we find that $\phi$ has a noticeable effect on the timing of interruptions.  We find that when $\theta=90^\circ$, close encounters with the secondary BHs can have noticeable effects on the dynamics of the stream.  The latest N-body simulations of axisymmetric nuclei reveal that the distribution of disrupted stars as a function of $\theta$ is double-peaked around $90^\circ$, due to stars on saucer orbits \citep{Zhong+2015}.  However, this depends in detail on the structure of the nuclear star cluster.  

\subsection{Accuracy of Light Curves}

There are several caveats relevant to the generation of our fallback rate curves.  First, the inclusion of hydrodynamics will alter the frequency of stream collisions and therefore the dynamical fate of infalling material.  Second, the rates we generate are sensitive to the particular value of the accretion radius.  One more caveat worth discussing further is that the existence of an accretion radius hinges on the assumption of an efficient circularisation mechanism.  Self-intersections due to apsidal precession can provide this mechanism, but this may only be efficient for the most massive BHs.  In addition, our code does not take into account nodal precession that occurs around rotating BHs.  SMBHs may have substantial spins, depending on their most recent accretion history \citep{Volonteri+2005,Volonteri+2013}.  Out-of-plane orbits around a rotating BH can therefore be substantially diverted, which can cause streams to miss each other and delaying or slowing accretion \citep{Dai+2013,Guillochon&Ramirez-Ruiz2015}.

Given these caveats, our fallback rates are most likely to translate to true light curves $L(t) \propto \dot{M}(t)$ when (i) the gaps in the fallback rate are small and (ii) nodal precession is negligible.  Large gaps in the fallback rate imply the existence of a large quantity of material in the system on chaotic orbits, which should experience extra collisions that are not accounted for in our simulations.  If nodal precession is important, which requires a rotating BH and a stellar orbit out of the plane, then accretion can be delayed or slowed, although this is important mostly for BHs of mass $\lesssim 10^6 \, \mathrm{M}_\odot$ \citep{Guillochon&Ramirez-Ruiz2015}.

\section{Conclusions}
\label{sec:conclusions}

We have performed simulations that trace the fallback of the debris of a TDE in the presence of a MBHB.  Compared to previous work, we have taken into account possible accretion onto the secondary BH, used an upgraded pseudo-Newtonian potential, performed experiments to determine when truncation occurs, explored how a MBHB affects the location of shocks, and scrutinised the key assumption that all material that enters $r_\mathrm{acc}$ is accreted.  The main results of this work are as follows:

\begin{itemize}
\item Gaps in the fallback rate appear when material misses the primary BH, typically because of the motion of the primary BH.  This missed material may or may not eventually accrete after a delay, even without accounting for hydrodynamics.  We expect that missed material will also be involved in shocks beyond the accretion radius, which are not captured in these simulations.
\item We distinguish between continuous and delayed accretion.  Continuous accretion originates from the infall of a coherent stream that persists throughout the end of the simulation, and follows the canonical power law, while delayed accretion originates from particles that initially miss the black hole and may eventually accrete randomly.  Continuous accretion is less likely to be affected by the addition of hydrodynamics than delayed accretion.
\item The secondary BH experiences the closest encounters with the stream during coplanar TDEs.  The secondary BH can imprint periodic signatures in the light curve due to these close encounters, and may accrete a significant amount of mass.  In extremely tight and equal-mass configurations, this can lead to stream trading, whereby the stream trades the BH that gets accreted onto.  The generation of powerful shocks around the secondary BH will alter our predictions for accretion.  These can only be resolved in hydrodynamic simulations. 
\item We find that the time of first interruption is a strong function of $\phi$.  The analytic estimate for truncation fails to take this dependence into account, and tends to break at extreme mass ratios.
\item We scrutinise the existence of an accretion radius.  We find that changing $r_\mathrm{acc}$ can dramatically alter inferred light curves.  Specifically, periods of delayed accretion can appear or vanish depending on the value of $r_\mathrm{acc}$ chosen.  
\item We find that unlike the case with a single BH, the distance of stream intersection varies significantly with time.  Shocks beyond the accretion radius may be important for enabling additional accretion onto the MBHB.
\end{itemize}

We hope that future hydrodynamical studies investigate the implications of these unusual shock conditions.  It would also be of interest to properly determine the amount of material that is accreted onto the secondary during close encounters with the stream.  Shocks that occur as matter is slingshot around the secondary BH may cause it to accrete additional matter.  

In cases when the two BHs accrete at comparable rates, it may be difficult to observationally disentangle the accretion onto each hole.  It is possible that the unusual accretion conditions caused by stream trading can lead to observable spectral variability.  In addition, if only one BH has high spin, it is possible that the triggering of a jet around that BH as in Sw 1644 \citep{Bloom+2011,Zauderer+2011,Levan+2015} can distinguish the two holes.  

Our results hint that light curves extracted from MBHB TDE simulations using ballistic orbits may look significantly different once hydrodynamics are taken into account.  Although we find gaps in the fallback rate due to the modified potential, as in L09 and L14, we also discover that the exact light curves we calculate depend critically on the value chosen for the accretion radius.  Consequently, while gaps in a TDE flare may indicate the presence of a MBHB, its parameters may be difficult to accurately extract without a better understanding of the hydrodynamics of the system.

\section*{acknowledgements} Angelo Ricarte performed this research with the support of the Gruber Science Fellowship.  PN acknowledges support from a NASA-NSF Theoretical and Computational Astrophysics Networks (TCAN) award number 1332858.  LD acknowledges NASA/NSF/TCAN (NNX14AB46G), NSF/XSEDE/TACC (TG-PHY120005), and NASA/Pleiades (SMD-14-5451).  We thank Enrico Ramirez-Ruiz for useful discussions.

\bibliography{ms}

\end{document}